\documentclass[fleqn,usenatbib]{mnras}

\usepackage{newtxtext,newtxmath}
\usepackage{multirow}

\usepackage[T1]{fontenc}

\DeclareRobustCommand{\VAN}[3]{#2}
\let\VANthebibliography\thebibliography
\def\thebibliography{\DeclareRobustCommand{\VAN}[3]{##3}\VANthebibliography}

\usepackage{graphicx}	
\usepackage{amsmath}	

\title[Polarimetric Properties of a Near--Sun Asteroid]{Polarimetric Properties of the Near--Sun Asteroid (155140) 2005 UD in Comparison with Other Asteroids and Meteoritic Samples}

\author[M. Ishiguro et al.]{
Masateru Ishiguro,$^{1,2} $ \thanks{E-mail: ishiguro@astro.snu.ac.kr (MIs)}
Yoonsoo P. Bach,$^{1,2} $
Jooyeon Geem,$^{1,2} $
Hiroyuki Naito,$^{3} $
Daisuke Kuroda,$^{4} $
\newauthor
Myungshin Im,$^{1,2} $
Myung Gyoon Lee,$^{1,2} $
Jinguk Seo,$^{1,2} $
Sunho Jin,$^{1,2} $
Yuna G. Kwon,$^{5} $
Tatsuharu Oono,$^{6} $
\newauthor
Seiko Takagi,$^{6} $
Mitsuteru Sato,$^{6} $
Kiyoshi Kuramoto,$^{6} $
Takashi Ito,$^{7} $
Sunao Hasegawa,$^{8} $
Fumi Yoshida,$^{9,10} $
\newauthor
Tomoko Arai,$^{10} $
Hiroshi Akitaya,$^{10} $
Tomohiko Sekiguchi,$^{11} $
Ryo Okazaki,$^{11} $
Masataka Imai,$^{12} $
Katsuhito Ohtsuka,$^{13} $
\newauthor
Makoto Watanabe,$^{14} $
Jun Takahashi,$^{15}$
Maxime Devog{\`e}le,$^{16} $
Grigori Fedorets,$^{17,18} $ 
Lauri Siltala,$^{17,19} $
\newauthor
and Mikael Granvik$^{17,20} $
\\
$^{1} $Department of Physics and Astronomy, Seoul National University, 1 Gwanak-ro, Gwanak-gu, Seoul 08826, Republic of Korea\\
$^{2} $SNU Astronomy Research Center, Seoul National University, 1 Gwanak-ro, Gwanak-gu, Seoul 08826, Republic of Korea\\
$^{3} $Nayoro Observatory, 157-1 Nisshin, Nayoro, Hokkaido 096-0066, Japan\\
$^{4} $Okayama Observatory, Kyoto University, 3037-5 Honjo, Kamogata, Asakuchi, Okayama 719-0232, Japan\\
$^{5} $Institut f\"{u}r Geophysik und Extraterrestrische Physik, Technische
Universit\"{a}t Braunschweig, Mendelssohnstr. 3, 38106 Braunschweig, Germany\\
$^{6} $Department of Cosmosciences, Graduate School of Science, Hokkaido University, Kita-ku, Sapporo, Hokkaido 060-0810, Japan\\
$^{7} $National Astronomical Observatory of Japan, 2-21-1 Osawa, Mitaka, Tokyo 181-8588, Japan\\
$^{8} $Institute of Space and Astronautical Science (ISAS), Japan Aerospace Exploration Agency (JAXA), Sagamihara, Kanagawa 252-5210, Japan\\
$^{9} $University of Occupational and Environmental Health, 1-1 Iseigaoka, Yahatanishi Ward, Kitakyushu, Fukuoka 807-8555, Japan\\
$^{10} $Planetary Exploration Research Center, Chiba Institute of Technology, Tsudanuma, Narashino, Chiba 275-0016, Japan\\
$^{11} $Asahikawa Campus, Hokkaido University of Education, Hokumon, Asahikawa, Hokkaido 070-8621, Japan\\
$^{12} $Faculty of Science, Kyoto Sangyo University,   Banyukan B401, Motoyama, Kamigamo, Kita-Ku, Kyoto-shi, Kyoto 603-8555, Japan\\
$^{13} $Tokyo Meteor Network, Daisawa 1-27-5, Setagaya, Tokyo 155-0032, Japan\\
$^{14} $Department of Applied Physics, Okayama University of Science, 1-1 Ridai-cho, Kita-ku, Okayama, Okayama 700-0005, Japan\\
$^{15} $Center for Astronomy, University of Hyogo, 407-2 Nishigaichi, Sayo, Hyogo 679-5313, Japan\\
$^{16} $Arecibo Observatory, University of Central Florida, HC-3 Box 53995, Arecibo, PR 00612, USA\\
$^{17} $Department of Physics, P.O. Box 64, FI-00014 University of Helsinki, Finland\\
$^{18} $Astrophysics Research Centre, School of Mathematics and Physics, Queen’s University Belfast, Belfast BT7 1NN, UK\\
$^{19} $Nordic Optical Telescope, Apartado 474, E-38700 S/C de La Palma, Santa Cruz de Tenerife, Spain\\
$^{20} $Asteroid Engineering Laboratory, Space Systems, Lule{\aa} University of Technology, Box 848, SE-98128 Kiruna, Sweden\\
}

\date{Accepted XXX. Received YYY; in original form ZZZ}

\newcommand{\hSlopeBoundLS}{\ensuremath{0.197}}
\newcommand{\dhSlopeBoundLShi}{\ensuremath{0.011}}
\newcommand{\dhSlopeBoundLSlo}{\ensuremath{0.012}}
\newcommand{\aZeroBoundLS}{\ensuremath{20.65}}
\newcommand{\daZeroBoundLShi}{\ensuremath{0.27}}
\newcommand{\daZeroBoundLSlo}{\ensuremath{0.29}}
\newcommand{\cOneBoundLS}{\ensuremath{1.159}}
\newcommand{\dcOneBoundLShi}{\ensuremath{0.063}}
\newcommand{\dcOneBoundLSlo}{\ensuremath{0.058}}
\newcommand{\cTwoBoundLS}{\ensuremath{0.000}}
\newcommand{\dcTwoBoundLShi}{\ensuremath{0.011}}
\newcommand{\dcTwoBoundLSlo}{\ensuremath{0.000}}
\newcommand{\aMinBoundLS}{\ensuremath{11.07}}
\newcommand{\daMinBoundLShi}{\ensuremath{0.23}}
\newcommand{\daMinBoundLSlo}{\ensuremath{0.23}}
\newcommand{\PMinBoundLS}{\ensuremath{-0.93}}
\newcommand{\dPMinBoundLShi}{\ensuremath{0.10}}
\newcommand{\dPMinBoundLSlo}{\ensuremath{0.09}}
\newcommand{\aMaxBoundLS}{\ensuremath{99.58}}
\newcommand{\daMaxBoundLShi}{\ensuremath{0.33}}
\newcommand{\daMaxBoundLSlo}{\ensuremath{0.36}}
\newcommand{\PMaxBoundLS}{\ensuremath{36.45}}
\newcommand{\dPMaxBoundLShi}{\ensuremath{1.06}}
\newcommand{\dPMaxBoundLSlo}{\ensuremath{1.01}}
\newcommand{\hSlopeBoundMC}{\ensuremath{0.197}}
\newcommand{\dhSlopeBoundMC}{\ensuremath{0.006}}
\newcommand{\aZeroBoundMC}{\ensuremath{20.65}}
\newcommand{\daZeroBoundMC}{\ensuremath{0.13}}
\newcommand{\cOneBoundMC}{\ensuremath{1.162}}
\newcommand{\dcOneBoundMC}{\ensuremath{0.029}}
\newcommand{\cTwoBoundMC}{\ensuremath{0.005}}
\newcommand{\dcTwoBoundMC}{\ensuremath{0.005}}
\newcommand{\aMinBoundMC}{\ensuremath{11.08}}
\newcommand{\daMinBoundMC}{\ensuremath{0.08}}
\newcommand{\aMaxBoundMC}{\ensuremath{99.53}}
\newcommand{\daMaxBoundMC}{\ensuremath{0.13}}
\newcommand{\PMinBoundMC}{\ensuremath{-0.93}}
\newcommand{\dPMinBoundMC}{\ensuremath{0.03}}
\newcommand{\PMaxBoundMC}{\ensuremath{36.43}}
\newcommand{\dPMaxBoundMC}{\ensuremath{0.37}}
\newcommand{\hSlopeUnboundLS}{\ensuremath{0.197}}
\newcommand{\dhSlopeUnboundLShi}{\ensuremath{0.007}}
\newcommand{\dhSlopeUnboundLSlo}{\ensuremath{0.008}}
\newcommand{\aZeroUnboundLS}{\ensuremath{19.71}}
\newcommand{\daZeroUnboundLShi}{\ensuremath{0.28}}
\newcommand{\daZeroUnboundLSlo}{\ensuremath{0.29}}
\newcommand{\cOneUnboundLS}{\ensuremath{0.734}}
\newcommand{\dcOneUnboundLShi}{\ensuremath{0.048}}
\newcommand{\dcOneUnboundLSlo}{\ensuremath{0.044}}
\newcommand{\cTwoUnboundLS}{\ensuremath{-1.894}}
\newcommand{\dcTwoUnboundLShi}{\ensuremath{0.174}}
\newcommand{\dcTwoUnboundLSlo}{\ensuremath{0.171}}
\newcommand{\aMinUnboundLS}{\ensuremath{8.44}}
\newcommand{\daMinUnboundLShi}{\ensuremath{0.34}}
\newcommand{\daMinUnboundLSlo}{\ensuremath{0.34}}
\newcommand{\PMinUnboundLS}{\ensuremath{-1.17}}
\newcommand{\dPMinUnboundLShi}{\ensuremath{0.08}}
\newcommand{\dPMinUnboundLSlo}{\ensuremath{0.08}}

\newcommand{\hSlopeUnboundMC}{\ensuremath{0.197}}
\newcommand{\dhSlopeUnboundMC}{\ensuremath{0.004}}
\newcommand{\aZeroUnboundMC}{\ensuremath{19.71}}
\newcommand{\daZeroUnboundMC}{\ensuremath{0.14}}
\newcommand{\cOneUnboundMC}{\ensuremath{0.735}}
\newcommand{\dcOneUnboundMC}{\ensuremath{0.022}}
\newcommand{\cTwoUnboundMC}{\ensuremath{-1.892}}
\newcommand{\dcTwoUnboundMC}{\ensuremath{0.081}}
\newcommand{\aMinUnboundMC}{\ensuremath{8.44}}
\newcommand{\daMinUnboundMC}{\ensuremath{0.13}}

\newcommand{\PMinUnboundMC}{\ensuremath{-1.17}}
\newcommand{\dPMinUnboundMC}{\ensuremath{0.03}}

\newcommand{\hSlopeUnboundLSSmallA}{\ensuremath{0.207}}
\newcommand{\dhSlopeUnboundLShiSmallA}{\ensuremath{0.011}}
\newcommand{\dhSlopeUnboundLSloSmallA}{\ensuremath{0.011}}
\newcommand{\aZeroUnboundLSSmallA}{\ensuremath{19.93}}
\newcommand{\daZeroUnboundLShiSmallA}{\ensuremath{0.33}}
\newcommand{\daZeroUnboundLSloSmallA}{\ensuremath{0.34}}
\newcommand{\cOneUnboundLSSmallA}{\ensuremath{0.801}}
\newcommand{\dcOneUnboundLShiSmallA}{\ensuremath{0.083}}
\newcommand{\dcOneUnboundLSloSmallA}{\ensuremath{0.077}}
\newcommand{\cTwoUnboundLSSmallA}{\ensuremath{-0.167}}
\newcommand{\dcTwoUnboundLShiSmallA}{\ensuremath{1.533}}
\newcommand{\dcTwoUnboundLSloSmallA}{\ensuremath{1.520}}
\newcommand{\aMinUnboundLSSmallA}{\ensuremath{8.89}}
\newcommand{\daMinUnboundLShiSmallA}{\ensuremath{0.52}}
\newcommand{\daMinUnboundLSloSmallA}{\ensuremath{0.54}}
\newcommand{\PMinUnboundLSSmallA}{\ensuremath{-1.20}}
\newcommand{\dPMinUnboundLShiSmallA}{\ensuremath{0.08}}
\newcommand{\dPMinUnboundLSloSmallA}{\ensuremath{0.08}}

\newcommand{\hSlopeUnboundMCSmallA}{\ensuremath{0.207}}
\newcommand{\dhSlopeUnboundMCSmallA}{\ensuremath{0.005}}
\newcommand{\aZeroUnboundMCSmallA}{\ensuremath{19.92}}
\newcommand{\daZeroUnboundMCSmallA}{\ensuremath{0.16}}
\newcommand{\cOneUnboundMCSmallA}{\ensuremath{0.803}}
\newcommand{\dcOneUnboundMCSmallA}{\ensuremath{0.037}}
\newcommand{\cTwoUnboundMCSmallA}{\ensuremath{-0.142}}
\newcommand{\dcTwoUnboundMCSmallA}{\ensuremath{0.714}}
\newcommand{\aMinUnboundMCSmallA}{\ensuremath{8.90}}
\newcommand{\daMinUnboundMCSmallA}{\ensuremath{0.20}}

\newcommand{\PMinUnboundMCSmallA}{\ensuremath{-1.20}}
\newcommand{\dPMinUnboundMCSmallA}{\ensuremath{0.03}}

\newcommand{\Deff}{\ensuremath{1.3}}
\newcommand{\DeffMax}{\ensuremath{1.38}}
\newcommand{\DeffMin}{\ensuremath{1.26}}

\newcommand{\pVMax}{\ensuremath{0.109}}
\newcommand{\pVMin}{\ensuremath{0.088}}

\pubyear{2021}

\begin{document}
\label{firstpage}
\pagerange{\pageref{firstpage}--\pageref{lastpage}}
\maketitle

\begin{abstract}
The investigation of asteroids near the Sun is important for understanding the final evolutionary stage of primitive solar system objects. 
A near-Sun asteroid, (155140) 2005 UD, has orbital elements similar to those of (3200) Phaethon (the target asteroid for the JAXA's {\it DESTINY$^+$} mission). We conducted photometric and polarimetric observations of 2005 UD and found that this asteroid exhibits a polarization phase curve similar to that of Phaethon over a wide range of observed solar phase angles ($ \alpha = 20 \mathrm{-} 105\degr $) but different from those of (101955) Bennu and (162173) Ryugu (asteroids composed of hydrated carbonaceous materials). At a low phase angle ($ \alpha \lesssim 30\degr$), the polarimetric properties of these near-Sun asteroids (2005 UD and Phaethon) are consistent with anhydrous carbonaceous chondrites, while the properties of Bennu are consistent with hydrous carbonaceous chondrites. We derived the geometric albedo, $ p_\mathrm{V} \sim 0.1 $ (in the range of $ \pVMin \mathrm{-} \pVMax $); mean $ V $-band absolute magnitude, $ H_\mathrm{V} = 17.54 \pm 0.02 $; synodic rotational period, $ T_\mathrm{rot} = 5.2388 \pm 0.0022 \,\mathrm{hours} $ (the two-peaked solution is assumed); and effective mean diameter, $ D_\mathrm{eff}  = 1.32 \pm 0.06 \,\mathrm{km} $. At large phase angles ($ \alpha \gtrsim 80\degr $), the polarization phase curve are likely explained by the dominance of large grains and the paucity of small micron-sized grains. We conclude that the polarimetric similarity of these near-Sun asteroids can be attributed to the intense solar heating of carbonaceous materials around their perihelia, where large anhydrous particles with small porosity could be produced by sintering.
\end{abstract}


\section{Introduction} \label{sec:introduction}
Among tens of thousands of known near-Earth asteroids (NEAs), asteroids with small perihelion distances (so-called near-Sun asteroids, NSAs, \citealt{2009PASJ...61.1375O,2013AJ....145..133J}) are attractive research targets in terms of the final evolutional stage of small solar system bodies. It was recently proposed that there could be catastrophic disruptions of NSAs at $ \lesssim$ 0.2 au from the Sun \citep{2016Natur.530..303G}, although the specific disruption mechanism is not clearly understood. (3200) Phaethon (formerly known as 1983 TB) is a typical NSA and was selected as the target of JAXA's {\it DESTINY$^+$} mission \citep{2018LPI....49.2570A}. Since its discovery in 1983, it has exhibited peculiar physical properties. It is dynamically linked to the Geminid meteor stream \citep{1983IAUC.3881....1W} and possibly other several streams \citep{2006A&A...450L..25O}). Phaethon has an asteroid-like orbit (i.e., the Tisserand parameter with respect to Jupiter's orbit, $T_\mathrm{J}<$3), but it exhibits weak recurrent activities like comets \citep{2013AJ....145..154L,2017AJ....153...23H}. Although such asteroid/comet-like hybrid objects have been discovered not only in the main asteroid belt \citep{2005ApJ...624.1093H} but also in near-Earth space \citep{2012AJ....143...66J}, Phaethon has another puzzling aspect of a dynamical association with (155140) 2005 UD, the target object of this study.

Table \ref{table:comparison} summarises the physical properties of these two NSAs, where the values written in boldface are obtained through our present work. \citet{2005CBET..283....1O} pointed out for the first time that 2005 UD indicated dynamical behaviour similar to Phaethon and suggested that 2005 UD could be a split nucleus of Phaethon \citep{2006A&A...450L..25O}. Later, \citet{2006AJ....132.1624J} conducted a photometric observation and supported the idea of \citet{2006A&A...450L..25O} because these two bodies have a bluish colour (B or F taxonomic type), which is rare among the small solar system bodies \citep[e.g.,][]{2004Icar..170..259B}. In the Tholen's taxonomy, B-types indicate a negative spectral slope (i.e., blue) with a moderate drop-off toward $0.4 \,\micron$, while F-types show a flat to slightly negative spectral slope with a weaker UV drop-off \citep{1984PhDT.........3T}.  \citet{2012Icar..218..196D} suggested B-types are further subcategorized into a wide variety of carbonaceous chondrite counterparts (from CM2 to CK4). Subsequently, \citet{2007A&A...466.1153K} noticed that the colour of 2005 UD changed with rotation, probably because of the surficial heterogeneity, and further speculated that the heterogeneity could result from fragmentation or collisional processes that occurred on the precursor of Phaethon and the 2005 UD. \citet{2019MNRAS.485.3378R} asserted that 2005 UD is not a member of the Phaethon--Geminid complex based on their dynamical analysis over the last 5,000 years; however, \citet{2016A&A...592A..34H}, and more recently \citet{2020arXiv201010633M}, suggested that the two objects might have separated from a common parent body a long time ago, approximately 10$^{5} $ years ago or, more likely, even before this epoch. On the contrary, \citet{2021arXiv210901020K} argued that the similar spectral property is only by coincidence from the analysis of their near-infrared spectrum.

\begin{table*}
\small
\centering
\caption{Comparison between Phaethon and 2005 UD}
\label{table:comparison} 
\begin{tabular}{lrr} 
\hline
& (3200) Phaethon & (155140) 2005 UD \\
\hline
Semimajor axis (au) & 1.271 & 1.275\\
Perihelion distance (au) & 0.140 & 0.163\\
Eccentricity & 0.890 & 0.872\\
Inclination (degree) & 22.26 & 28.67\\
Tisserand parameter with respect to Jupiter & 4.510 & 4.507\\

Synodic rotational period (hr) & 3.6039 (0.0004)$^{a} $ & 5.249$^{j} $, 5.231$^{k} $, 5.235 (0.005)$^l$ \\
& &5.237 (0.001)$^m$ {\bf 5.2388 (0.0022)}\\
Sidereal rotational period (hr) & 3.6039$^{a,b,c} $ & 5.2340 ($_{-0.00001}^{+0.00004} $)$^n$ \\
Spectral type & B, F, C & B, F, C $^{j, k, l} $ \\
Absolute magnitude in $V$-band & 14.24$^{d} $, 14.27 (0.04)$^{b} $, &17.48 (0.04)$^{j*} $, 17.51 (0.02)$^l$,\\
& 13.63 (0.02)$^{e} $ & {\bf 17.54 (0.02)}\\
Geometric albedo & 0.122 (0.008)$^{b,c} $, 0.14 (0.04)$^{f} $, & 0.14 (0.09)$^{h} $, 0.10 (0.02)$^l$,\\
& 0.08 (0.01)$^{g} $, 0.16 (0.02)$^{h} $ &  $ \mathbf{\pVMin \mathrm{-} \pVMax} $ \\
Diameter (km) & 4.6 ($^{+0.2}_{-0.3} $)$^{h} $, 5.1 (0.2)$^{b,c} $, $>$6.0 $^i$, 5.4 (0.5)$^l$ & 1.2 (0.4)$^{h} $, 1.3 (0.1)$^{j,l} $, $ \mathbf{\DeffMin \mathrm{-} \DeffMax} $ \\
\hline
\multicolumn{3}{l}{$^a$ \citet{2018A&A...619A.123K},
$^b$ \citet{2016A&A...592A..34H},
$^c$ \citet{2018A&A...620L...8H},
$^d$ \citet{2014ApJ...793...50A},
$^e$ \citet{2019AJ....158...30T},}\\
\multicolumn{3}{l}{$^f$ \citet{2018ApJ...864L..33S},
$^g$ \citet{2018AJ....156..287K},
$^h$ \citet{2019AJ....158...97M},
$^i$ \citet{2019P&SS..167....1T},
$^j$ \citet{2006AJ....132.1624J},
}\\
\multicolumn{3}{l}{$^k$ \citet{2007A&A...466.1153K},
$^l$ \citet{2020PSJ.....1...15D},
$^m$ \citet{2019EPSC...13.1989K},
$^n$ \citet{2021P&SS..19505120H}.}\\
\multicolumn{3}{l}{The errors are shown in parentheses. The values written in boldface were obtained through this work.}\\
\multicolumn{3}{l}{The orbital elements were obtained from the JPL Small-Body Database Browser (\url{https://ssd.jpl.nasa.gov/sbdb.cgi\#top}). }\\
\end{tabular}
\end{table*}

Polarimetric studies on Phaethon were recently conducted, and different research groups published a series of papers. First, \citet{2018NatCo...9.2486I} noticed through their 2016 observations that Phaethon exhibited a large polarization degree of up to $ \sim 50 \,\% $ at the largest phase angle (Sun--asteroid--observer angle) of their observation ($ \alpha = 106.5\degr $). 
Later, \citet{2018ApJ...864L..33S} derived the geometric albedo of $ p_\mathrm{V} = 0.14 \pm 0.04 $ via the polarimetric slope and geometric albedo law and found that the geometric albedo is significantly larger than the comet nuclei \citep{2004Icar..167...16B,2009Icar..204..209L,2013Icar..222..467L,2013Icar..226.1138F,2014ApJ...789..151K,2015A&A...583A..31C}. \citet{2018MNRAS.479.3498D} conducted independent polarimetric observations in 2017 and noticed that Phaethon's polarimetric inversion angle, $ \alpha_0 $ (the phase angle when the polarization degree is zero) was within the range of typical asteroids but beyond the range of F-type asteroids and cometary nuclei, therefore supporting the idea of asteroidal origin. \citet{2018MNRAS.480L.131B} utilised a set of data in \citet{2018MNRAS.479.3498D} and further found that the rotational variation in the polarization degree was probably caused by local heterogeneity. \citet{2018ApJ...864L..33S} and \citet{2020P&SS..18004774O} pointed out that the polarization degree of Phaethon in 2017 was different from that in 2016 at larger phase angles ($ \alpha > 60$\degr) and conjectured that Phaethon might have large-scale surficial inhomogeneity.

We conducted the polarimetric observation using the same instruments as \citet{2018NatCo...9.2486I} employed for Phaethon observation, which provides a reliable comparison between these two NSAs. Moreover, we re-analysed polarimetric data acquired through observations in \citet{2020PSJ.....1...15D}, including a set of unpublished data at a large phase angle. We also made a photometric observation at the opposition ($ \alpha \sim 1 \degr $) for deriving the absolute magnitude and diameter. In Section 2, we describe our observations and data analysis. We report our findings in Section 3. In Section 4, we provide an interpretation of our polarimetric results compared to other asteroids and meteorite samples.

\section{Observations and Data Analysis} \label{sec:observation}

\begin{table*}
\scriptsize
\caption{Observation Circumstance}
\label{table:obs}
\begin{tabular}{cccccccccccc}
\hline
Date & UT & Telescopes/Instruments & Mode$^a$ & Filter & Exptime$^b$ & $N^c$ & Airmass & $ r^d $ & $ \Delta^e $ & $ \alpha^f $ & $ \phi^g $ \\
   &  &  &  & & (sec) & & & ($ \mathrm{au} $) & ($ \mathrm{au} $) & (deg) & (deg)\\
\hline
 2018 Sep 24 &  15:07--16:14 &  NO/MSI &  Pol &  $R_\mathrm{C}$ &    180 &    12 &  1.94--2.96 &  1.07 &   0.23 &  68.08 &  269.41 \\
 2018 Sep 25 &  15:59--19:38 &  NO/MSI &  Pol &  $R_\mathrm{C}$ &    120 &    28 &  1.26--1.90 &  1.08 &   0.23 &  63.60 &  269.16 \\
 2018 Sep 27 &  14:29--15:36 &  NO/MSI &  Pol &  $R_\mathrm{C}$ &     90 &    24 &  1.75--2.51 &  1.11 &   0.22 &  55.44 &  268.55 \\
 2018 Oct 02 &  13:04--19:45 &  NO/MSI &  Pol &  $R_\mathrm{C}$ &     60 &  128 &  1.24--2.22 &  1.19 &   0.23 &  33.49 &  266.37 \\
 2018 Oct 03 &  13:32--19:20 &  NO/MSI &  Pol &  $R_\mathrm{C}$ &     60 &  260 &  1.24--1.78 &  1.20 &   0.24 &  29.28 &  265.93 \\
 2018 Oct 04 &  11:55--19:42 &  NO/MSI &  Pol &  $R_\mathrm{C}$ &     60 &  328 &  1.24--2.75 &  1.21 &   0.24 &  25.47 &  265.57 \\
 2018 Oct 08 &  13:36--17:04 &  NO/MSI &  Pol &  $R_\mathrm{C}$ &     90 &  100 &  1.25--1.40 &  1.27 &   0.28 &  11.67 &  265.11 \\
 2018 Oct 09 &  13:11--18:46 &  NO/MSI &  Pol &  $R_\mathrm{C}$ &     90 &  172 &  1.25--2.07 &  1.28 &   0.29 &   8.66 &  265.69 \\
\\
2018 Oct 12 & 10:53--19:46 & SAO/STX-16803 & Photo & $R_\mathrm{C}$ & 60 & 413 & 1.16--2.53 & 1.32 & 0.32 &1.14 &265.68 \\
2018 Oct 13 & 10:36--18:57 & SAO/STX-16803 & Photo & $R_\mathrm{C}$ & 60 & 398 & 1.16--2.59 & 1.33 & 0.34 &1.35 &265.68 \\
\\
2018 Sep 12 & 05:21--05:29 &  NOT/ALFOSC+FAPOL & Pol &  $R_\mathrm{C}$ & 120 & 4 & 2.14--2.27 & 0.86 & 0.34 & 106.47 & 266.93 \\
2018 Sep 19 & 04:58-05:18 & NOT/ALFOSC+FAPOL & Pol &  $R_\mathrm{C}$ & 90 & 8  & 1.40--1.51 & 0.98 & 0.27 & 87.74 & 269.52 \\
2018 Sep 30 & 05:03--05:34 &  NOT/ALFOSC+FAPOL & Pol &  $R_\mathrm{C}$ & 60 & 12 & 1.08--1.11 & 1.15 & 0.22 & 44.01 & 267.47 \\
2018 Oct 01 & 05:49-05:53 & NOT/ALFOSC+FAPOL & Pol &  $R_\mathrm{C}$ & 60 & 4  & 1.17--1.17 & 1.17 & 0.23 & 39.53 & 267.00 \\
2018 Oct 02 &02:47--03:39 &  NOT/ALFOSC+FAPOL & Pol &  $R_\mathrm{C}$ & 60 & 40 & 1.07--1.11 & 1.18 & 0.23 & 35.71 & 266.60 \\
2018 Oct 04 &23:57--00:44 &  NOT/ALFOSC+FAPOL & Pol &  $R_\mathrm{C}$ & 60 & 28 & 1.27--1.48 & 1.22 & 0.24 & 24.08 & 265.44 \\
2018 Oct 05 & 22:54--05:32 &  NOT/ALFOSC+FAPOL & Pol &  $R_\mathrm{C}$ & 60 & 208 & 1.07--1.85 & 1.23 & 0.25 & 20.15 & 265.17 \\
2018 Oct 11 & 23:22--00:16 &  NOT/ALFOSC+FAPOL & Pol &  $R_\mathrm{C}$ & 60 & 32 & 1.10--1.20 & 1.31 & 0.31 & 2.58 & 274.48 \\
2018 Oct 12 & 21:02-21:38 & NOT/ALFOSC+FAPOL & Pol &  $R_\mathrm{C}$ & 60 & 32 & 1.63--2.00 & 1.32 & 0.33 & 0.74 & 317.11 \\
2018 Oct 14 & 01:36-01:53 & NOT/ALFOSC+FAPOL & Pol &  $R_\mathrm{C}$ & 60 & 16 & 1.10-1.12 & 1.34 & 0.34 & 2.25 & 63.71 \\
2018 Oct 15 & 00:57-01:58 & NOT/ALFOSC+FAPOL & Pol &  $R_\mathrm{C}$ & 60 & 36 & 1.08-1.15 & 1.35 & 0.35 & 4.18 & 70.82 \\
2018 Oct 17 & 02:10-03:03 & NOT/ALFOSC+FAPOL & Pol &  $R_\mathrm{C}$ & 75 & 36 & 1.22-1.43 & 1.38 & 0.38 & 7.83 & 74.11 \\

\hline
\multicolumn{11}{l}{$^a $ Observation mode (Photo: photometry, Pol: polarimetry), $ ^b $ Exposure time in seconds. $ ^c $ Number of valid exposures, $ ^d $ Median heliocentric distance in $ \mathrm{au} $}\\
\multicolumn{11}{l}{$ ^e $ Median geocentric distance in $ \mathrm{au} $, $ ^f $ Median solar phase angle in degrees, $ ^g $ Position angle of the scattering plane in degrees.}\\
\multicolumn{11}{l}{The web-based JPL Horizon system (\url{http://ssd.jpl.nasa.gov/?horizons}) was used to obtain $r$, $ \Delta$, $ \alpha$, and $ \phi$ in the table.}\\
\end{tabular}
\end{table*}

\subsection{Observations}
\label{subsec:observation}
Table \ref{table:obs} shows the summary of our observations. We performed polarimetric observations for 9 nights from 2018 September 24 to 2018 October 09 using the 1.6-m Pirka Telescope at the Nayoro Observatory of Faculty of Science, Hokkaido University (NO) , Japan (Minor Planet Center observatory code Q33). We employed a Multi-Spectral Imager (MSI) mounted at the $ f/12 $ Cassegrain focus of the telescope \citep{2012SPIE.8446E..2OW}. In the standard imaging mode, MSI covers a field-of-view (FOV) of $3.3\arcmin \times 3.3\arcmin $  with $ 0.39 \arcsec \mathrm{pixel}^{-1}$ resolution. MSI has an imaging polarization mode covering two adjacent sky areas of $ 3.3\arcmin\ \times 0.7\arcmin$ each which are separated by $ 1.7\arcmin $ with a polarization mask. We conducted the imaging polarimetry using the southern part of the sky in the FOVs, inserting the polarization mask, Wollaston beam splitter, and rotatable $ \lambda /2 $ plate into the MSI optical path. We chose the standard $ R_\mathrm{C} $-band filter (with the central wavelength at $ 0.64 \,\micron $ and the effective bandwidth of $ 0.15 \,\micron $, see, \citealt{2012SPIE.8446E..2OW}). We operated the telescope mount in asteroid tracking mode, so background objects (e.g., stars and galaxies) were trailed in the FOV. During the observations, we examined the signal-to-noise ratio (S/N) and tuned individual exposure times in the range from 60 to 180 seconds to archive $ \mathrm{S/N} \sim 10 \mathrm{-} 100 $ in the single exposures. At the beginning of the polarimetric run ($ \alpha \geq 46.41 \degr $), we could not obtain substantial numbers of polarimetric images ($ N $ in Table \ref{table:obs}) because of unfavourable weather conditions. However, we obtained sufficient numbers of images after October 2 ($ \alpha \leq 33.50\degr $) owing to clear-sky conditions.

In addition to the above polarimetry, we conducted photometric observations for 2 nights on 2018 October 12 and 13 using the 1-m telescope at the Seoul National University Astronomical Observatory (SAO) on the Gwanak campus, Seoul, South Korea \citep{IM:2021}. Although the observatory is located at the southern edge of a large metropolitan area where the sky is severely affected by light pollution, the specifications of the telescope and instruments are sufficient to obtain meaningful lightcurve data for the bright asteroid ($ \sim 15.7 \,\mathrm{mag} $) on these nights. The observations were performed taking advantage of the rare observation opportunity when the asteroid was located in the opposite direction from the Sun (i.e., the solar phase angle $ \alpha = 0.8\degr \mathrm{-} 1.5\degr $). Accordingly, the observation data offer a forte for enabling derivation of the absolute magnitude (defined as a magnitude observed at opposition from the unit observer's and heliocentric distances). We utilised the Santa Barbara Instrument Group (SBIG) STX-16803 CCD camera ($ 4096 \times 4096 $ pixels at $ 9 \,\micron$) mounted on the $ f/6 $ Nasmyth focus. This combination of the telescope and CCD camera covers the FOV of $ 21.1\arcmin \times 21.1\arcmin$ with a pixel scale of $ 0.31 \arcsec \mathrm{pixel}^{-1} $. The telescope was operated in asteroid tracking mode.

Table \ref{table:obs} also contains information on data acquired using the 2.5-m Nordic Optical Telescope (NOT, MPC code Z23) at the Observatorio del Roque de los Muchachos, La Palma, Canary Islands, Spain. These data were reanalysed in this work. The description of the observation is given in \citet{2020PSJ.....1...15D}. The data were acquired with the ALFOSC instrument and the FAPOL polarimeter. A broadband filter called \texttt{R\_Bes 650\_130} was used for the observation. Since the transmittance of the filter is very similar to that of the MSI $ R_\mathrm{C} $-band filter, we regard \texttt{R\_Bes 650\_130} as the standard $ R_\mathrm{C} $-band filter in this paper. The polarimetric images were acquired through a calcite plate and a rotatable $ \lambda$/2 plate. Because a field mask was not inserted for the observations, the ordinary and the extraordinary components are overlapped together with a small offset angle ($ 15\arcsec $). The combination of these instruments covers a circular FOV of $ \sim 1\arcmin $ with $ 0.43\arcsec \mathrm{pixel}^{-1} $ resolution. It is important to notice that there is a set of polarimetric images taken at a very large phase angle ($ 106.47 \degr $) but not published in \citet{2020PSJ.....1...15D} because only one set of polarimetric images was acquired on that night (UT  2018 September 12). Because these data are important to constrain the maximum polarization degree ($P_\mathrm{max}$), we analysed the data with great care, as shown below.

\subsection{Data Analysis}
\label{subsec:analysis}

We analysed the MSI polarimetric data in the same manner as \citet{2017AJ....154..180I} and \citet{2018NatCo...9.2486I}. The outline consists of (i) preprocessing using bias and dome--flat data, (ii) cosmic-ray rejection, (iii) masking field stars near the target asteroid, (iv) source flux extraction from ordinary and extraordinary regions on MSI images using the aperture photometry algorithm while avoiding the masked regions for the field stars, and (v) derivation of the Stokes parameters ($ I $, $ Q $, and $ U $), the linear polarization degree ($ P $), and the position angle of polarization ($ \theta_\mathrm{P} $). Since the details about the reduction and error analyses are given in these reference papers, we do not repeat the description in this paper. However, there is one difference regarding the process step (v) that is worth explaining. Since the primary and secondary mirrors of the Pirka Telescope were cleaned on 2017 February 11 (i.e., after the Phaethon observation and before the 2005 UD observation), it was thought that the cleaning process might have changed the instrumental polarization parameters. We obtained the polarimetric calibration data in 2018 March and June--2019 September to examine the secular change after the maintenance period (see Table \ref{table:polcal}). Over three years, the change in calibration parameters created only a $0.024 \mathrm{-} 0.086$\% difference in the polarization degree, comparable to or even smaller than the weighted mean errors of our final polarimetric results. Although the difference was small, we applied the set of parameters obtained in 2018 March (after the maintenance) to provide reliable data sets.

\begin{table*}
\scriptsize
\caption{Polarimetric Calibration Parameters ($R_{\rm C} $-band)}
\label{table:polcal}
\begin{tabular}{lcccccl}
\hline
Date & Instruments & $ P_\mathrm{eff} $$^{*1} $ & $q_\mathrm{inst} $$^{*2} $ & $u_\mathrm{inst} $$^{*3} $ & $ \theta_\mathrm{inst} $$^{*4} $ & Remarks\\
\hline
2016 Oct & MSI & $ 99.48 \pm 0.03 $ & $0.705 \pm 0.017 $ & $ 0.315 \pm 0.016 $ & $ 3.94 \pm 0.31 $ & Calibration data in \citet{2018NatCo...9.2486I}\\
&&&&&&UP$^{*5} $: G191B2B, HD21447 \\
&&&&&&SP$^{*6} $: HD19820 ($114.46\pm0.16$), HD25443 ($133.65\pm0.28$)\\
2018 Mar & MSI & $99.13 \pm 0.01 $ & $ 0.791 \pm 0.025 $ & $ 0.339 \pm 0.020 $ & $ 3.66 \pm 0.17 $ & Calibration data in this study\\
&&&&&&UP$^{*5} $:G191B2B, GD319, Gamma Boo, HD154892, HD21447\\
&&&&&&SP$^{*6} $: HD19820 ($114.46\pm0.16$), HD204827 ($59.10\pm0.17$),\\ 
&&&&&& HD25443 ($133.65\pm0.28$)\\
2019 Jun--Sep & MSI & $99.60 \pm 0.01 $ & $ 0.828 \pm 0.006 $ & $ 0.311 \pm 0.006 $ & $ 3.17 \pm 0.07 $ &--\\
&&&&&&UP$^{*5} $: HD14069, HD154892, HD212311\\
&&&&&&SP$^{*6} $: BD+64d106 ($96.74\pm0.54$), HD155197 ($102.88\pm0.18$),\\
&&&&&& HD161056 ($67.33\pm0.23$), HD204827 ($59.10\pm0.17$),\\
&&&&&&Hiltner 960 ($54.54\pm0.16$)\\

2018 Sep--Oct & NOT/ALFOSC+FAPOL & $ 100 $ (assumed) &$  -0.043 \pm 0.065 $ & $ -0.077 \pm 0.075 $ & $ 93.10 \pm 0.06 $ & Calibration data in this study \\
&&&&&&UP$^{*5} $: BD+28 4211, BD+32 3739, G191B2B, HD14069 \\
&&&&&&SP$^{*6} $: BD+59 389 ($98.14\pm0.1$), VI Cyg $\#$12 ($116.23\pm0.14$)\\
\hline
\multicolumn{6}{l}{$^{*1} $ Polarimetric efficiency in \%, see, \citealt{2017AJ....154..180I},
$^{*2} $ Instrumental polarization of $Q/I$ in \%, $^{*3} $ Instrumental polarization $U/I$ in \%,}\\
\multicolumn{6}{l}{$^{*4} $ Reference position angle of the polarization in degrees.}\\
\multicolumn{6}{l}{$^{*5} $ Unpolarized standard stars. We regarded these polarization degrees as zero.}\\
\multicolumn{6}{l}{$^{*6} $ Polarized standard stars. The catalogued position angles in degrees are given in the parentheses.}\\
\end{tabular}
%

\end{table*}

We analysed the ALFOSC/FAPOL polarimetric data in a manner similar to the MSI data. The instrumental polarization parameters are examined by observing polarimetric standard stars (Table \ref{table:polcal}). We paid particular attention to field stars in this polarization data analysis. Because signals from ordinary and extraordinary components overlap in the obtained images, the asteroid signal is occasionally contaminated by the field stars. 
In particular, the asteroid most frequently encountered field stars on 2018 September 19 because it was located close to the galactic plane (the galactic latitude of 1.5 \degr). We contrived a technique to eliminate field stars (see Appendix \ref{app: NOT}). By this process, field stars vanished from the sky region near the asteroid in most images, making it possible to derive the polarization degree for the night. For the data of 2018 September 12 ($ \alpha =106.47\degr $), we found that there are no field stars brighter than 20.8 magnitudes (i.e., stars listed in the Gaia catalogue, \citealt{2018A&A...616A...1G}) passing within the aperture of the asteroid. We also checked whether there are hot pixels and cosmic rays within the aperture of the asteroid and found no such pixels. For these reasons, we derived the polarization degree at the largest phase angle  ($ \alpha = 106.47\degr $) even from the single set of polarimetric images. Because the weighted mean is not available for the data on 2018 September 12, we append the error based on the S/N and the systematic error associated with the instrumental polarization parameters.

The photometric data were preprocessed in the standard manner for CCD data. The original object images were bias- and dark-subtracted and flat-fielded using the dome flat. The cosmic ray was then removed using the L. A. Cosmic algorithm \citep{2001PASP..113.1420V} implemented in \texttt{astroscrappy}\footnote{\url{https://github.com/astropy/astroscrappy} version 1.0.8 with a separable median filter and specifically tuned parameters.}. Then, the World Coordinate System (WCS) information was appended in each image header by the offline version of \texttt{astrometry.net} \citep{2010AJ....139.1782L}.
We queried the Pan-STARRS1 DR1 (hereafter DR1) catalogue \citep{2020ApJS..251....7F} $ r $ magnitude after preprocessing in the range between 10.0 and 15.2 mag, and toggled flags if there were DR1-catalogued objects near the target asteroid in order to avoid contamination of the photometric signal of the asteroid. In addition, we discarded the extracted objects from the DR1 catalogue if any pairs of stars were close to them. We only selected objects that were (1) not recognised as a quasar, galaxy, or variable star based on the DR1 catalogue's flags and (2) observed several instances in the shorter wavelengths (at least three times for the $g$- and $r$-band and once for the $i$-band). Finally, we had a minimum of 5 to a maximum of more than 20 stars in each image. The magnitudes of these stars were used for the photometric calibration, as explained below.

The aperture shape of each star is designed as that of a pill-box. It is a combination of a rectangle and two half-ellipses, similar to \texttt{TRIPPy} \citep{2016AJ....151..158F}. The position angle of the aperture is obtained by fitting the two-dimensional elliptical Gaussian to field stars with the initial guess from the ephemerides, and the sigma-clipped median of the angles of the field stars is used to determine the aperture position angles. After testing many combinations of the semi-major and minor axis lengths of the half-ellipses, we empirically determined the appropriate solution of half-circles with a radius $ 1.75F $, where $ F $ is the full-width at half-maximum (FWHM) of the point sources, to enclose a sufficient amount of the stellar signal even when the tracking accuracy of the telescope mount was not perfect. Therefore, the aperture was set as a combination of a rectangle with width $ L $, the expected trail length of the asteroid concerning the field stars during the exposure time retrieved from the ephemerides, and height $ 3.50F $, with two half-circles with radius $ 1.75F $. The instrumental magnitude of each frame was calculated by subtracting the sky value estimated from the locally defined pill-box annulus with inner and outer radii of $ 4F $ and $ 6F $, respectively, while retaining the same width ($L$) of the rectangle. For the asteroid, which is the tracked target, we set a circular aperture with a radius of $2F$ and the circular annulus for the sky flux with inner and outer radii of $ 4F $ and $ 6F $. We confirmed that a change of the apertures' sizes affected the results only within much less than the estimated 1-$ \sigma$ error bars.

The magnitudes in DR1 were converted into Johnson--Cousins $R_\mathrm{C} $ magnitudes by the transformation formula given in \cite{2012ApJ...750...99T}. By comparing the instrumental and catalogue magnitudes, we determined the photometric zero point of each image. The instrumental magnitudes were then converted to the standardised magnitudes using the photometric zero point. We ignored the colour-term for the atmospheric extinction, which would be negligible for stars of $ 0 \lesssim g-r \lesssim 1 $ from our analysis (zero point slope $ \lesssim 0.05 $ for $ g-r $).

The observed $ R_\mathrm{C} $ magnitudes, $ R $, were converted into reduced magnitudes (hypothetical magnitudes at the unit heliocentric distance of 1 au and the observer's distance of 1 au), which is given by,
\begin{equation} \label{eq:eq1}
  m_\mathrm{R}(1,1,\alpha)=R - 5\log_{10}(r_\mathrm{h} \Delta),
\end{equation}
where $ r_\mathrm{h} $ and $ \Delta $ are the heliocentric and the observer's distances in au during the epoch of our observation. Since our photometric data were acquired at the opposition (i.e., $ \alpha \sim 1\degr $), we ignored the $ \alpha $-dependency of the magnitude and derived the absolute magnitude $ H_\mathrm{R} := m_\mathrm{R}(1,1,0) $ in the $ R_\mathrm{C} $-band.

To obtain the lightcurve, we further corrected the light time to consider the asteroid's rotation while light travelled to the observer's location. Finally, we manually inspected each image with the locations of photometric apertures to check whether our photometric results were affected by unexpected problems (such as close encounters with background objects that are not listed in the DR1 or imperfect centering of objects due to the low S/N, and so on.). Of 810 images, 41 images were excluded due to such unexpected situations. Since data points with a large scatter (29 data points) were automatically rejected in the period analysis, 740 photometric data points were used in this work.

\section{Results}
\label{sec:results}

In this section, we report our polarimetric and photometric findings separately as below.

\subsection{Phase Angle Dependence of Polarization Degree}\label{subsec:polresult}

\begin{figure*}
 \includegraphics[width=1.5\columnwidth]{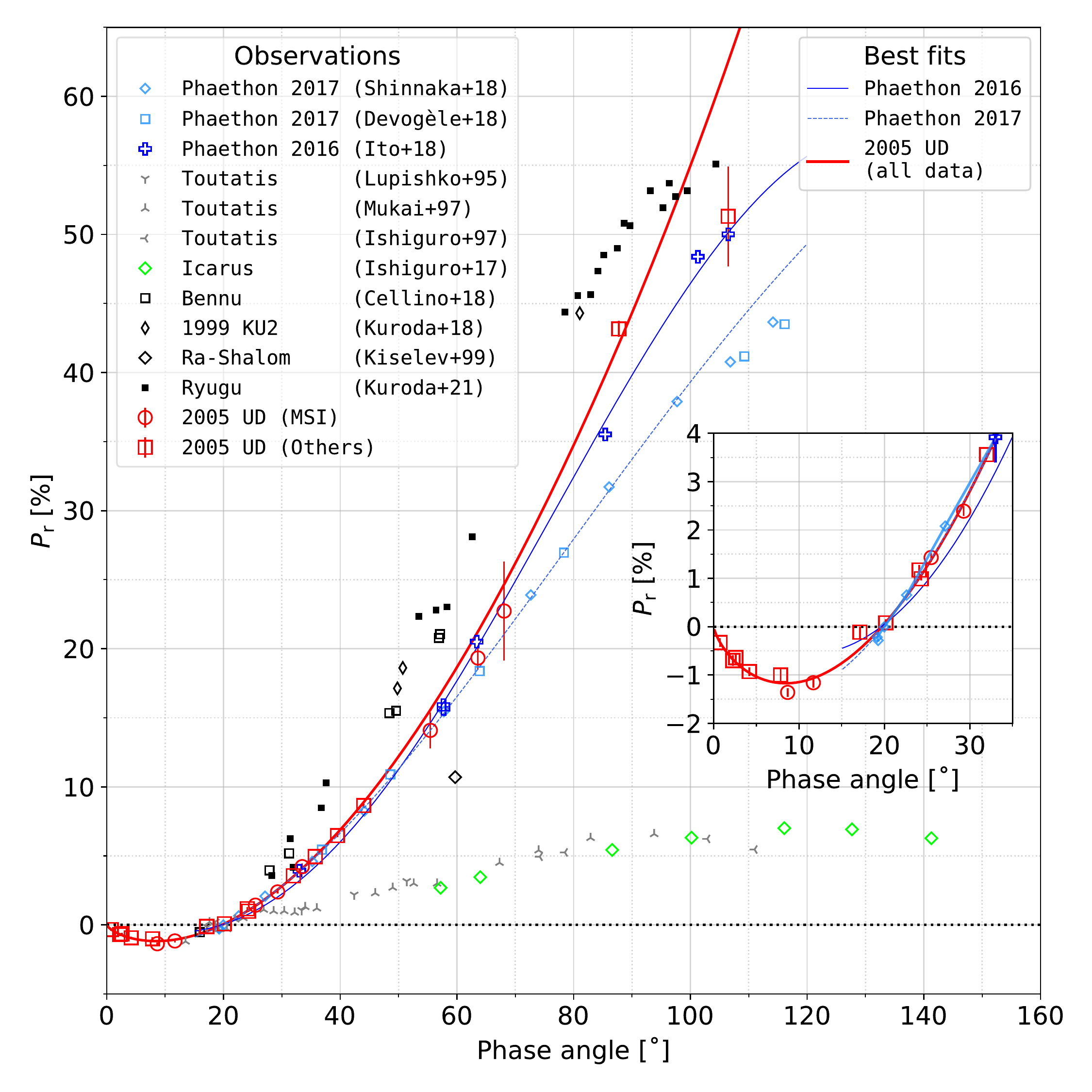}
\caption{Phase angle ($ \alpha$) dependence of polarization degree ($ P_\mathrm{r} $). We show the data for the 2005 UD together with Phaethon and S-type ((4179) Toutatis and (1566) Icarus) and C-type ((101955) Bennu, (152679) 1998 KU$_2$ and (2100) Ra-Shalom) asteroids for comparison. The polarization phase curves of Phaethon observed in 2016 and 2017 were separately fitted by a trigonometric function. We show fitted curves of Phaethon only at $\alpha > 15\degr$ because the fitting result in the negative branch looks strange because of the lack of data points. The references for comparison include \citet{2018ApJ...864L..33S}, \citet{2018MNRAS.479.3498D}, \citet{2018NatCo...9.2486I}, \citet{1995Icar..113..200L}, \citet{1997Icar..127..452M}, \citet{1997PASJ...49L..31I}, \citet{2017AJ....154..180I}, \citet{2018MNRAS.481L..49C}, \citet{2018A&A...611A..31K}, \citet{2021ApJ...911L..24K}, and \citet{1999Icar..140..464K}.}
\label{fig:fig1}
\end{figure*}

The weighted mean values of the nightly polarimetric data are given in Table \ref{table:polvalues}. We also show the phase angle dependence of polarization degrees in Figure \ref{fig:fig1}. The data cover a wide area of the solar phase angles up to $ \alpha = 106.47\degr $. In Figure \ref{fig:fig1}, we show the polarization degrees of Phaethon and several asteroids (C- and S-groups, which are common in the near-Earth space). At a glance, it is evident that 2005 UD exhibits a polarization phase curve consistent with that of Phaethon but significantly different from that of S-group asteroids, as already noticed in \citet{2020PSJ.....1...15D}. Moreover, the polarization phase curve of 2005 UD at lower phase angles ($ \alpha \lesssim 60\degr $) is not as steep as those of C--type asteroids, (101955) Bennu \citep{2018MNRAS.481L..49C}, (152679) 1998 KU$_2$ \citep{2018A&A...611A..31K}, and (162173) Ryugu \citep{2021ApJ...911L..24K}.
Because the polarization slope around the inversion angle is primarily dependent on the albedos but less dependent on particle sizes \citep{1986MNRAS.218...75G}, it is reasonable to hypothesize that 2005 UD and Phaethon have similar albedo values for the observed wavelength ($R_\mathrm{C} $-band).

\begin{table*}
\caption{Polarimetric Results}
\label{table:polvalues}
\begin{tabular}{ccccccccc}
\hline
Date& $ \alpha$ & $ P^a$ & ${\sigma P}^b$ & $ \theta_\mathrm{P}^c $ & ${\sigma\theta_\mathrm{P}}^d $ & ${P_\mathrm{r}}^e $ &  ${\theta_\mathrm{r}}^f$ & Telescopes/Instruments\\
& & ($ \% $) & ($ \% $) & (deg) & (deg) & ($ \% $) & (deg) \\
\hline

2018 Sep 24 & $ 68.08 $ & $ 22.74 $ & $ 3.59  $ & $ 0.33 $ & $ 4.52 $ & $ 22.73 $ & $    0.91 $  & NO/MSI \\
2018 Sep 25 & $ 63.60 $ & $ 19.45 $ & $ 0.84  $ & $ -4.02 $ & $ 1.24 $ & $ 19.33 $ & $   -3.18 $  & NO/MSI \\
2018 Sep 27 & $ 55.44 $ & $ 14.10 $ & $ 1.29  $ & $ -0.80 $ & $ 2.63 $ & $ 14.09 $ & $   0.65 $  & NO/MSI \\
2018 Oct 02 & $ 33.49 $ & $ 4.22  $ & $ 0.22  $ & $ -5.38 $ & $ 1.53 $ & $  4.22 $ & $   -1.75 $  & NO/MSI \\
2018 Oct 03 & $ 29.28 $ & $ 2.40  $ & $ 0.10  $ & $ -5.37 $ & $ 1.21 $ & $  2.39 $ & $   -1.31 $  & NO/MSI \\
2018 Oct 04 & $ 25.47 $ & $ 1.43  $ & $ 0.09  $ & $ -4.63 $ & $ 1.86 $ & $  1.43 $ & $    -0.20 $  & NO/MSI \\
2018 Oct 08 & $ 11.67 $ & $ 1.16  $ & $ 0.10  $ & $ 82.57 $ & $ 2.41 $ & $ -1.16 $ & $ 87.46 $  & NO/MSI \\
2018 Oct 09 & $ 8.66  $ & $ 1.37  $ & $ 0.09  $ & $ 82.81 $ & $ 1.79 $ & $ -1.36 $ & $  87.13 $  & NO/MSI \\ \\


 2018 Sep 12 & $ 106.47 $ & $ 51.44 $ &  $ 3.62 $ &  $ -5.11 $ &   $  2.01 $ &  $ 51.31 $ &   $ -2.04 $ & NOT/ALFOSC+FAPOL\\
 2018 Sep 19 &  $ 87.74 $ &  $ 43.18 $ & $ 0.59 $ &  $ -1.24 $  &    $ 0.39 $ &  $ 43.16 $ &   $  -0.76 $ &NOT/ALFOSC+FAPOL\\
 2018 Sep 30 &  $ 44.01 $ &  $ 8.68 $  & $ 0.14 $ & $ -5.24 $  &   $  0.47 $ &  $ 8.65 $ &    $ -2.72 $ &NOT/ALFOSC+FAPOL\\
 2018 Oct 01 & $ 39.53 $ &  $ 6.49 $  &  $ 0.13 $ &  $ -5.42 $  &    $ 0.56 $ &  $ 6.47 $ &    $ -2.43 $ &NOT/ALFOSC+FAPOL\\
 2018 Oct 02 &  $ 35.71 $ &  $ 4.97 $  & $ 0.10 $ &  $ -6.05 $   &    $ 0.60 $ & $ 4.95 $ &    $ -2.64 $ &NOT/ALFOSC+FAPOL\\
 2018 Oct 04 &  $ 24.08 $ &  $ 1.17 $  & $ 0.11 $ & $ -3.46 $   &    $ 2.60 $ &  $ 1.17 $ &     $ 1.10 $ &NOT/ALFOSC+FAPOL\\
 2018 Oct 05 &  $ 20.15 $ &  $ 0.08 $  & $ 0.07 $ & $ 3.63 $   &    $ 24.76 $ &  $ 0.08 $ &    $ 8.49 $ &NOT/ALFOSC+FAPOL\\
 2018 Oct 11 &   $ 2.58 $ &  $ 0.64 $  &  $ 0.08 $ &  $ -88.20 $  &    $ 3.77 $ &  $ -0.64 $ &    $ 87.32 $ &NOT/ALFOSC+FAPOL\\
 2018 Oct 12 &   $ 0.74 $ &  $ 0.38 $  &  $ 0.09 $ & $ -28.21 $  &    $ 7.05 $ &  $ -0.33 $ &    $ 104.65 $ & NOT/ALFOSC+FAPOL\\
 2018 Oct 14 &   $ 2.25 $ &  $ 0.84 $  & $ 0.11 $ &  $ 80.46 $  &    $ 3.80 $ &  $ -0.70 $ &    $ 106.74 $ & NOT/ALFOSC+FAPOL\\
 2018 Oct 15 &   $ 4.18 $ &   $ 0.93 $  & $ 0.09 $ &  $ 72.25 $  &    $ 2.81 $ &  $ -0.93 $ &    $ 91.43 $ &NOT/ALFOSC+FAPOL\\
 2018 Oct 17 &   $ 7.83 $ &  $ 1.02 $  &  $ 0.12 $ &  $ 68.38 $  &    $ 3.50 $ &  $ -1.00 $ &    $ 84.28 $ &NOT/ALFOSC+FAPOL\\

\hline         
\multicolumn{8}{l}{$ ^a $Nightly averaged polarization degree as a percentage.}\\
\multicolumn{8}{l}{$ ^b $Uncertainty of $ P$ as a percentage.}\\
\multicolumn{8}{l}{$ ^c $ Position angle of the strongest electric vector in degrees.}\\
\multicolumn{8}{l}{$ ^d $Uncertainty of $ \theta_\mathrm{P} $ in degrees. }\\
\multicolumn{8}{l}{$ ^e $ Polarization degree referring to the scattering plane as a percentage. It is defined as $P_\mathrm{r}=P \cos\left(2\theta_\mathrm{r}\right)$
}\\
\multicolumn{8}{l}{$ ^f $ Position angle referring to the scattering plane in degrees. It is given as $\theta_\mathrm{r}=\theta_\mathrm{P}-(\phi\pm90\degr)$}\\
\end{tabular}
\end{table*}

We fit the polarization phase curve using an empirical function that has been widely employed for the $ P_\mathrm{r}(\alpha) $ curves \citep{1993LPICo.810..194L}:
\begin{equation}\label{eq:LM}
  P_\mathrm{r}(\alpha) = 
    h 
    \left( \frac{\sin \alpha}{\sin \alpha_0} \right)^{c_1} 
    \left(\frac{\cos \frac{\alpha}{2}}{\cos \frac{\alpha_0}{2}}\right)^{c_2} 
    \sin(\alpha - \alpha_0) ~,
\end{equation}
where $ h $, $ c_1 $, $ c_2 $, and $ \alpha_0 $ are all free parameters for fitting the $ P_\mathrm{r}(\alpha) $ curve. In Eq. (\ref{eq:LM}), we modified the original formula so that $h$ coincided with the slope at the polarimetric inversion angle $ \alpha = \alpha_0 $. This empirical formula was probably contrived because $ P_\mathrm{r} = 0 $ is guaranteed at $ \alpha = 0\degr $, $ \alpha_0 $, and $ 180\degr $ when $ c_1 > 0 $ and $ c_2 > 0 $. However, the restriction on $ c_1 $ and $ c_2 $ sometimes prevents us from fitting some phase curves. Thus, recent polarimetric observations of NSA, (1566) Icarus, over a very large $\alpha$ range suggest a limitation of Eq. (\ref{eq:LM}) that $ P_\mathrm{r}(\alpha) $ data cannot be expressed by this equation when the restriction of $ c_1 > 0 $ and $ c_2 > 0 $ is applied \citep{2017AJ....154..180I}. Therefore, we also tested the data fitting without the restriction. Hereafter, we call the former case ($ c_1 > 0 $ and $ c_2 > 0 $) the ``bound'' case and the latter the ``unbound'' case. Moreover, we fitted the data at small phase angles ($\alpha<45\degr$). We assumed $ \alpha_0 \in [10\degr,\, 35\degr] $ and $ h \in [0\,\mathrm{\% \,deg}^{-1},\, 1\,\mathrm{\%/ \,deg}^{-1}] $ for both cases.

\begin{figure*}
 \includegraphics[width=2.0\columnwidth]{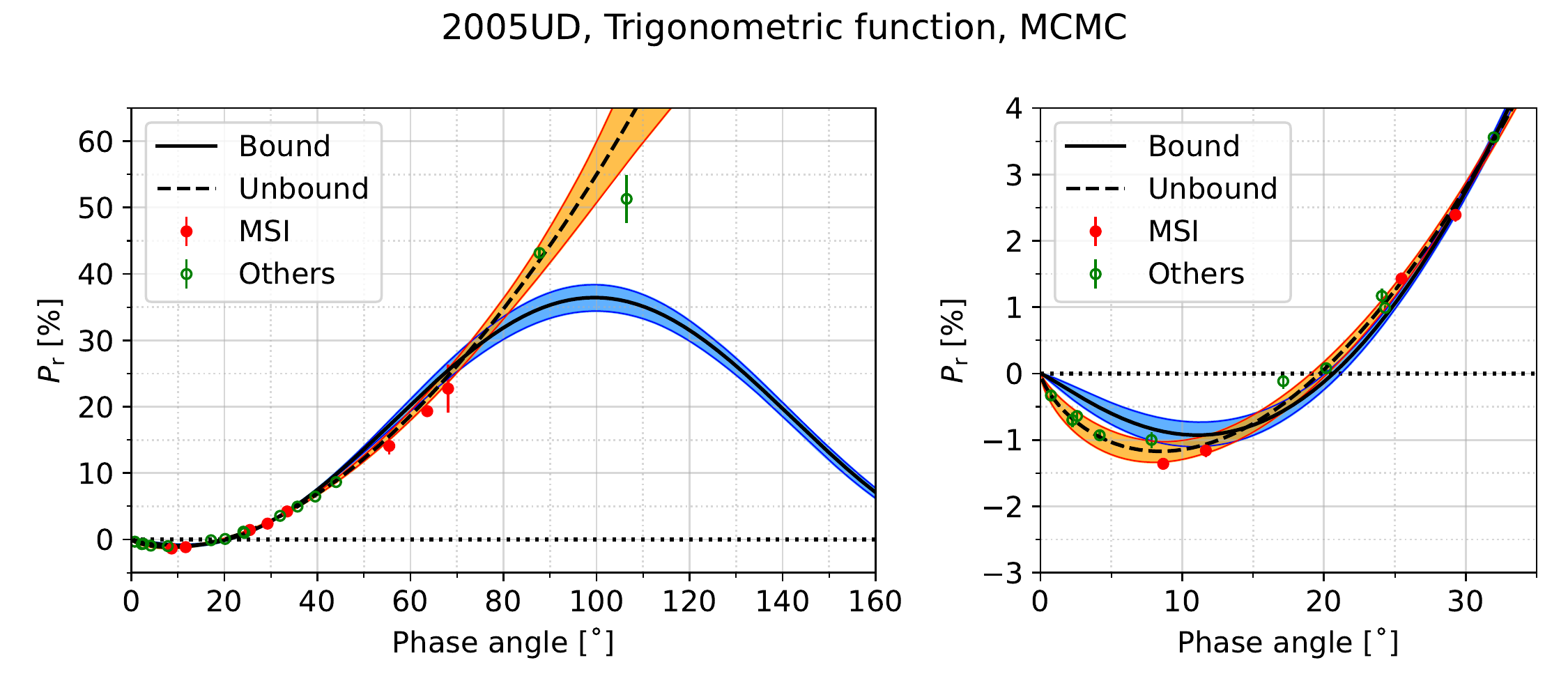}
  \caption{The observed data acquired by MSI (red filled circle), ALFOSC, and FoReRo2 (green open circle) overplotted with best-fit functions using Eq. (\ref{eq:LM}) for bound (solid) and unbound (dashed) cases. The shades indicate the uncertainty of the curve fittings based on MC simulation (blue and orange for bound and unbound cases, respectively).}
  \label{fig:bestfit}
\end{figure*}

The detailed descriptions of the fitting and the code are given in Appendix \ref{app: fitting}. We employed the Markov chain Monte Carlo (MCMC) method implemented in PyMC3 \citep{2016ascl.soft10016S} and standard least-square (minimum $ \chi^2 $) root finding to perform a comprehensive search for the best-fit parameters and their uncertainties.
We compiled data in Table \ref{table:polvalues} and the data acquired with FoReRo2 \citep{2020PSJ.....1...15D} for the fitting. The fitting results are summarised in Table \ref{table:polfits} and shown in Figure \ref{fig:bestfit}. 
Two parameters, $ \alpha_\mathrm{max} $ and $ P_\mathrm{max} $, are not well determined because the polarization phase curve keeps increasing even at the largest observed phase angle ($ \alpha = 106.47\degr $). When $ c_2 > 0 $, the polarization phase curve was not fitted to the data at large phase angles (Figure \ref{fig:bestfit} left). The bound case also does not work in the small phase angles (see the blue area in Figure \ref{fig:bestfit}). 
Therefore, there are discrepancies in $P_\mathrm{min}$ and $\alpha_\mathrm{min}$ between these three cases. 
However, we obtained a set of consistent and reliable results of $h$, $\alpha_0$, $\alpha_\mathrm{min}$, and $P_\mathrm{min}$  for both the unbound (all) and unbound ($\alpha <45\degr$) cases. Almost all of the observed data points at low phase angles are in good agreement with the model curve within the margin of error (see the orange area in Figure \ref{fig:bestfit} right). In the following discussion, we adopted the results of polarimetric parameters obtained in the unbound case for both all data and $\alpha < 45\degr$ data. 

\begin{table*}
  \centering
  \caption{Polarimetric Fitting Results}
    \begin{tabular}{c|l|rrrrrrrr}
    \hline
    \multicolumn{1}{c}{\multirow{2}[1]{*}{Boundness} $ ^a $} & 
    \multicolumn{1}{c}{\multirow{2}[1]{*}{Results} $ ^b $} & 
    \multicolumn{1}{c}{$ h $} & \multicolumn{1}{c}{$ \alpha_0 $} & 
    \multicolumn{1}{c}{$ c_1 $} & \multicolumn{1}{c}{$ c_2 $} & 
    \multicolumn{1}{c}{$ \alpha_\mathrm{min} $} & 
    \multicolumn{1}{c}{$ P_\mathrm{min} $} & 
    \multicolumn{1}{c}{$ \alpha_\mathrm{max} ^c $} & 
    \multicolumn{1}{c}{$ P_\mathrm{max} ^c $} 
    \\
    \multicolumn{1}{c}{} &
    & 
    \multicolumn{1}{c}{($ \% \,\mathrm{deg}^{-1} $)} & 
    \multicolumn{1}{c}{($ \degr $)} & 
    \multicolumn{1}{c}{-} &
    \multicolumn{1}{c}{-} & 
    \multicolumn{1}{c}{($ \degr $)} & 
    \multicolumn{1}{c}{($ \% $)} & \multicolumn{1}{c}{($ \degr $)} & 
    \multicolumn{1}{c}{($ \% $)} 
    \\
    \hline
      \multirow{5}[4]{*}{Bound (all data)} 
        & LS  &         \hSlopeBoundLS    & \aZeroBoundLS    & \cOneBoundLS    & \cTwoBoundLS    & \aMinBoundLS    & \PMinBoundLS    & (\aMaxBoundLS)    & (\PMaxBoundLS) \\
              & $ + $ & \dhSlopeBoundLShi & \daZeroBoundLShi & \dcOneBoundLShi & \dcTwoBoundLShi & \daMinBoundLShi & \dPMinBoundLShi & (\daMaxBoundLShi) & (\dPMaxBoundLShi) \\
              & $ - $ & \dhSlopeBoundLSlo & \daZeroBoundLSlo & \dcOneBoundLSlo & \dcTwoBoundLSlo & \daMinBoundLSlo & \dPMinBoundLSlo & (\daMaxBoundLSlo) & (\dPMaxBoundLSlo) \\
        \cline{2-10}
              & MC    & \hSlopeBoundMC  & \aZeroBoundMC  & \cOneBoundMC  & \cTwoBoundMC  & \aMinBoundMC  & \PMinBoundMC  & (\aMaxBoundMC)  & (\PMaxBoundMC) \\
              & sd    & \dhSlopeBoundMC & \daZeroBoundMC & \dcOneBoundMC & \dcTwoBoundMC & \daMinBoundMC & \dPMinBoundMC & (\daMaxBoundMC) & (\dPMaxBoundMC) \\
        \cline{1-10}
      \multirow{5}[3]{*}{Unbound (all data)} 
        & LS  &         \hSlopeUnboundLS    & \aZeroUnboundLS    & \cOneUnboundLS    & \cTwoUnboundLS    & \aMinUnboundLS    & \PMinUnboundLS    & - & - \\
              & $ + $ & \dhSlopeUnboundLShi & \daZeroUnboundLShi & \dcOneUnboundLShi & \dcTwoUnboundLShi & \daMinUnboundLShi & \dPMinUnboundLShi & - & - \\
              & $ - $ & \dhSlopeUnboundLSlo & \daZeroUnboundLSlo & \dcOneUnboundLSlo & \dcTwoUnboundLSlo & \daMinUnboundLSlo & \dPMinUnboundLSlo & - & - \\
        \cline{2-10}
              & MC    & \hSlopeUnboundMC  & \aZeroUnboundMC  & \cOneUnboundMC  & \cTwoUnboundMC  & \aMinUnboundMC  & \PMinUnboundMC  & - & - \\
              & sd    & \dhSlopeUnboundMC & \daZeroUnboundMC & \dcOneUnboundMC & \dcTwoUnboundMC & \daMinUnboundMC & \dPMinUnboundMC & - & - \\
    \hline
    \multirow{5}[3]{*}{Unbound ($ \alpha < 45^\circ $)} 
        & LS  &         \hSlopeUnboundLSSmallA    & \aZeroUnboundLSSmallA    & \cOneUnboundLSSmallA    & \cTwoUnboundLSSmallA    & \aMinUnboundLSSmallA    & \PMinUnboundLSSmallA    & - & - \\
              & $ + $ & \dhSlopeUnboundLShiSmallA & \daZeroUnboundLShiSmallA & \dcOneUnboundLShiSmallA & \dcTwoUnboundLShiSmallA & \daMinUnboundLShiSmallA & \dPMinUnboundLShiSmallA & - & - \\
              & $ - $ & \dhSlopeUnboundLSloSmallA & \daZeroUnboundLSloSmallA & \dcOneUnboundLSloSmallA & \dcTwoUnboundLSloSmallA & \daMinUnboundLSloSmallA & \dPMinUnboundLSloSmallA & - & - \\
        \cline{2-10}
              & MC    & \hSlopeUnboundMCSmallA  & \aZeroUnboundMCSmallA  & \cOneUnboundMCSmallA  & \cTwoUnboundMCSmallA  & \aMinUnboundMCSmallA  & \PMinUnboundMCSmallA  & - & - \\
              & sd    & \dhSlopeUnboundMCSmallA & \daZeroUnboundMCSmallA & \dcOneUnboundMCSmallA & \dcTwoUnboundMCSmallA & \daMinUnboundMCSmallA & \dPMinUnboundMCSmallA & - & - \\
    \hline
  \multicolumn{10}{l}{$ ^a $``Bound'' and ``Unbound'' indicate conditions if $ c_1 ,\, c_2 > 0 $ is considered (the former) or not (the latter).}\\
  \multicolumn{10}{l}{$ ^b $``LS'': least-square solution, ``$+/-$'': 1-$ \sigma$ uncertainty bounds from the least-square method using MC samples,}\\
    \multicolumn{10}{l}{~~~``MC'' and ``sd'': the mean and standard deviation of the Monte Carlo samples, respectively.}\\
    \multicolumn{10}{l}{$ ^c $ $ \alpha_\mathrm{max} $ and $ P_\mathrm{max} $ are less reliable and thus are in parentheses or omitted (see Section \ref{subsec:polresult} and Appendix \ref{app: fitting}).}
     \end{tabular}%
  \label{table:polfits}%
\end{table*}

\subsection{Geometric Albedo}
It is known that the polarization slope $ h $ exhibits good correlation with the geometric albedo (the so-called polarimetric slope--albedo law). The correlation was noticed by \citet{1967AnWiD..27..109W} and \citet{1967JGR....72.3105K} for the first time. The relation is understandable phenomenologically when considering that multiple scattering between individual constitutive scattering elements randomises the scattering plane so that a highly reflective surface tends to indicate a small polarization degree and, therefore, a low polarization slope \citep{1971A&A....12..199D}. The polarimetric slope--albedo law is written as
\begin{equation} \label{eq:slopealbedo}
  \log_{10}(p_\mathrm{V}) = C_1 \log_{10}(h) + C_2 ~~,
\end{equation}
where $ C_1 $ and $C_2 $ are constants. The uncertainty is obtained by
\begin{equation} 
\label{eq:slopealbedo_error}
\Delta p_\mathrm{V} 
  \approx p_\mathrm{V} \ln 10 
    \sqrt{ 
      (\log_{10} (h) \Delta C_1)^2 
      + (\Delta C_2)^2
      + \left( \frac{C_1 \Delta h}{h \ln 10} \right)^2
    } ~.
\end{equation}

\noindent In Eq. (\ref{eq:slopealbedo_error}), we take account of the error of the polarization slope ($\Delta h$) and the errors of these coefficients ($ \Delta C_1$ and $ \Delta C_2$). These constants and errors have been examined using different sets of observational data. We employed these values from the latest publications \citep{2015MNRAS.451.3473C, 2018SoSyR..52...98L}, and we obtained the geometric albedo in the $ R_\mathrm{C} $-band of $ p_\mathrm{R} \approx 0.1 $ (Table \ref{table:polalbedo}). It is important to notice that the geometric albedo is usually defined in the $ V $-band rather than the $ R_\mathrm{C} $-band. We obtained the polarimetric data using the $ R_\mathrm{C} $-band filter because the Pirka/MSI band provides more reliable data (i.e., smaller error) than the $ V $-band. In this paper, we regard $ p_\mathrm{V} = p_\mathrm{R} $ in the following discussion because the colour index $ (V-R_\mathrm{C}) = 0.35 \pm 0.02 $ for 2005 UD \citep{2006AJ....132.1624J} effectively matches $(V-R_\mathrm{C})_\odot = 0.354 \pm 0.010 $ for the Sun \citep{2006MNRAS.367..449H}, suggesting that the albedo values are less dependent on wavelength between these bands (i.e., $V$ and $R_\mathrm{C} $).


\begin{table*}
\centering
\caption{Geometric albedo values derived using different sets of $ C_1 $ and $C_2 $}
\label{table:polalbedo} 
\begin{tabular}{l c c c c c} 
\hline
\multicolumn{3}{c}{} & \multicolumn{3}{c}{Geometric albedo$^{*1} $ } \\
& $ C_1 $ & $ C_2 $ & bound (all) & unbound (all) & unbound ($\alpha<45\degr)$\\
\hline
\citet{2015MNRAS.451.3473C} & $ -1.111 \pm 0.031 $ & $ -1.781 \pm 0.025 $ &
    $ 0.101 \pm 0.008 $ & $ 0.101 \pm 0.008 $ & $ 0.096 \pm 0.008 $\\
\citet{2018SoSyR..52...98L} & $ -1.016 \pm 0.010 $ & $ -1.719 \pm 0.012 $ &
    $ 0.099 \pm 0.004 $ & $ 0.099 \pm 0.004 $ & $ 0.095 \pm 0.004$ \\
\hline
\multicolumn{5}{l}{$^{*1} $  $ p_\mathrm{R} = p_\mathrm{V} $ is assumed.}\\
\end{tabular}
\end{table*}

\subsection{Photometric Result and 2005 UD's Diameter}\label{subsec:photoresult}
\begin{figure}
\includegraphics[width=\columnwidth]{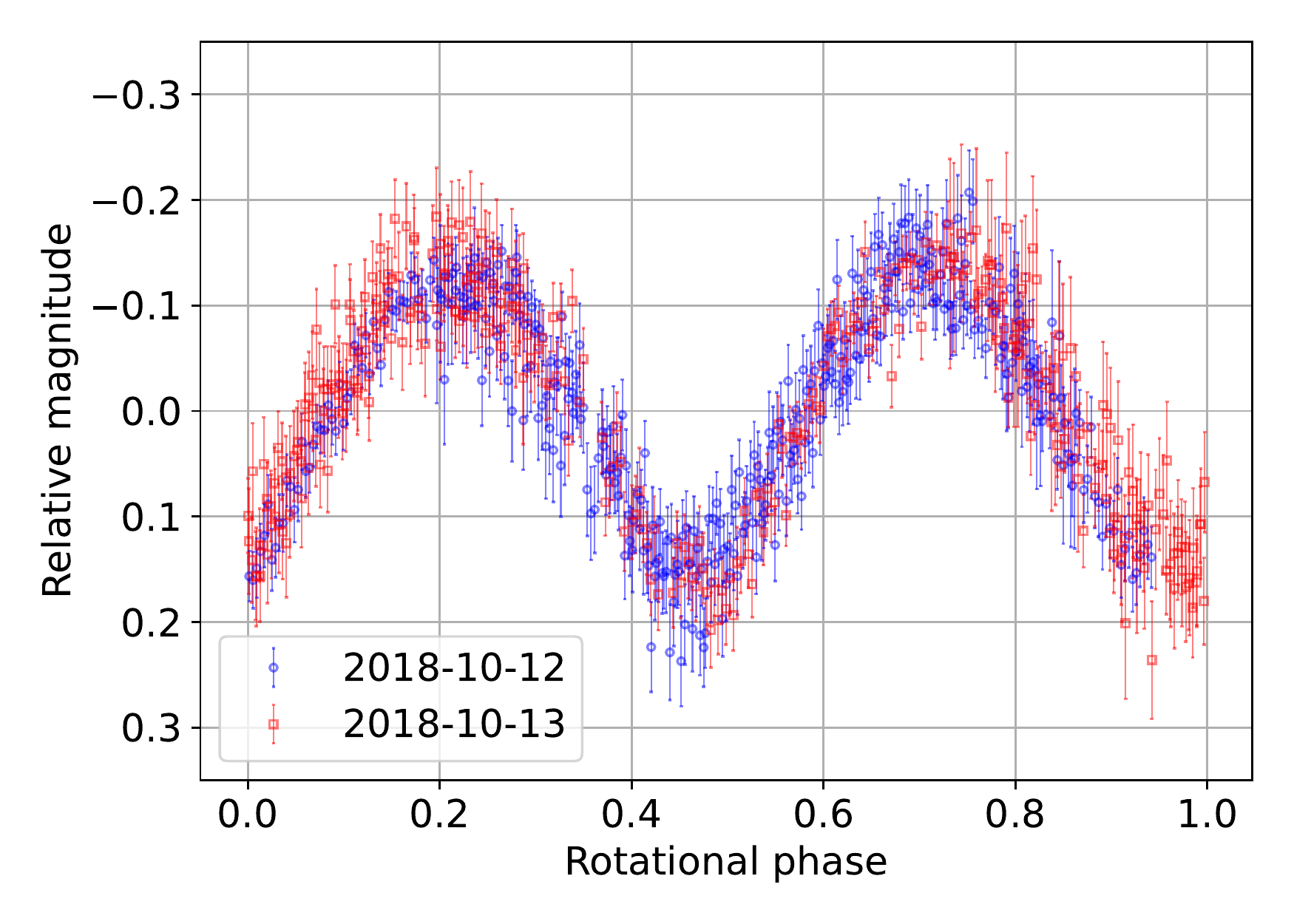}
\caption{The lightcurve folded with the rotational period of 5.2388 hours. Of 810 measurements, 740 data points were used (see Section \ref{subsec:photoresult}).
}
\label{fig:lightcurve}
\end{figure}

Figure \ref{fig:lightcurve} shows the lightcurve at the phase angle $ \alpha = 0.8\degr\mathrm{-}1.5\degr $. After correction of the distance effect using Eq. (\ref{eq:eq1}), we obtained the reduced magnitudes near the opposition, which were almost equivalent to the absolute magnitude $ H_\mathrm{R} $ because of the small phase angle. We obtained the mean absolute magnitudes of 17.182 on 2018 October 12 and 17.189 on 2018 October 13 in the $ R_\mathrm{C} $-band. We utilised the generalised Lomb-Scargle periodogram \citep{2009A&A...496..577Z} to determine the synodic rotational period and obtained $ T_\mathrm{rot} = 5.2388 \pm 0.0022 $ hours, assuming that one rotation creates two peaks and two troughs. For confirmation, we constructed the lightcurve folded with the determined $T_\mathrm{rot} $ and confirmed that the lightcurve data obtained at different times were effectively overplotted (Figure \ref{fig:lightcurve}). Since the shape and amplitude of the lightcurve are similar for these peaks and troughs, we cannot rule out the cases for more than three peaks and troughs in one rotation as a solution. However, as discussed in \citet{2020PSJ.....1...15D} that more than three peaked lightcurve is less likely, we adopt a rotation period of $ T_\mathrm{rot} = 5.2388 \pm 0.0022 $ hours. From the lightcurve, we derived an amplitude of 0.293 mag, which corresponds to the apparent axis ratio of 1.31.

The effective diameter $D$ (km) is given by the following equation:
\begin{equation} 
\label{eq:sizealbedo}
  D = \frac{C}{\sqrt{p_\mathrm{V}}} 10^{-H_\mathrm{V}/5} ~,
\end{equation}
where $C = 2 \,\mathrm{au} \times 10^{V_\odot/5} = 1329 \mathrm \,\mathrm{km} $ is a constant ($V_\odot$ is the $V$-band magnitude of the Sun at 1 au; \citealt{2007Icar..190..250P}). It should be noted that the constant $C$ was derived for the absolute magnitude $H_\mathrm{V} $ rather than $H_\mathrm{R} $. Considering again that the colour index of the asteroid $(V - R_\mathrm{C}) = 0.35 \pm 0.02 $ \citep{2006AJ....132.1624J}, we obtained the mean $V$-band absolute magnitude of 2005 UD as $H_\mathrm{V} = 17.54 \pm 0.02 $. Substituting $H_\mathrm{V} $ and $ p_\mathrm{V} $ into Eq. (\ref{eq:sizealbedo}), we found the apparent diameter of $ D \sim \Deff \,\mathrm{km} $. Strictly, the diameter ranges from $ D = \DeffMin \,\mathrm{km} $ for $ p_\mathrm{V} = \pVMax $ to $ D = \DeffMax\,\mathrm{km} $ for $ p_\mathrm{V} = \pVMin $ using the unbound case in Table \ref{table:polalbedo}. The minimum and maximum values of $ p_\mathrm{V} $ are calculated according to the lower bound of the minimum and upper bound of the maximum albedo estimation in the unbound case, excluding the last row in the table.

\section{Discussion}
\label{sec:discussion}

The derived albedo, diameter, rotational period, and absolute magnitude were compared with previous results (see Table \ref{table:comparison}). All of them are consistent with each other, strengthening the reliability of these results. In the following subsection, we compare our polarimetric results with those of other asteroids and laboratory samples, and we conjecture a corresponding meteorite type and surface physical condition (porosity and grain size). 

\subsection{\texorpdfstring{$ \alpha_0 $}{a0}--\texorpdfstring{$ P_\mathrm{min} $}{Pmin} relation}
\label{subsection:aP}
\begin{figure*}
\includegraphics[width=1.8\columnwidth]{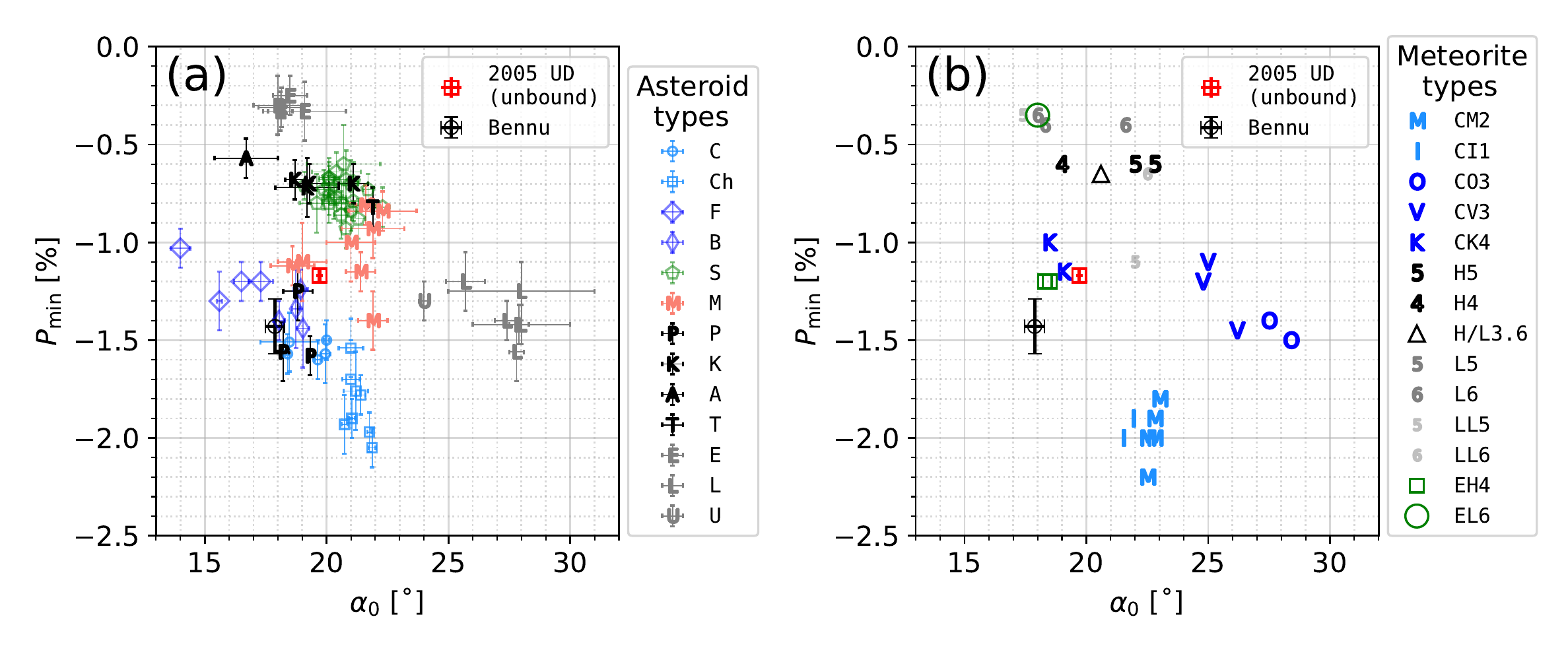}\\
\caption{Comparison of $ \alpha_0 $--$ P_\mathrm{min} $ between 2005 UD and (a) other asteroids and (b) meteorites. We chose 2005 UD results based on the fitting of all data in the unbound case (see Table \ref{table:polfits}). }
\label{fig:compasteroids}
\end{figure*}

To begin with, we examine two parameters ($ P_\mathrm{min} $ and $ \alpha_0 $) for characterizing the negative branch of the polarization phase curve: $ P_\mathrm{min} $ is the minimum polarization degree, and $ \alpha_0 $ is the inversion angle at which $ P_\mathrm{r} (\alpha_0) = 0 $ takes place. $ \alpha_0 $ is sometimes notated as $ \alpha_\mathrm{inv} $ in some literature. Figure \ref{fig:compasteroids} (a) indicates the comparison of the $ \alpha_0 $--$ P_\mathrm{min} $ relation between 2005 UD and other asteroids. These $ \alpha_0 $--$ P_\mathrm{min} $ data of asteroids other than 2005 UD and Bennu are given in Figure 5 of \citet{2017Icar..284...30B}. In the Belskaya database, taxonomic types were appended using information obtained by either \citet{1984PhDT.........3T} or \citet{2009Icar..202..160D}. It is known that $ P_\mathrm{min} $ depends not only on albedo, but also on surficial texture (i.e., porosity and grain size, \citealt{1975LPSC....6.2749D}), while $ \alpha_0 $ is more sensitive to surficial texture \citep[e.g., the existence of subwavelength small grains,][]{1990MNRAS.245...46G}. 
However, as mentioned in \citet{2017Icar..284...30B}, asteroids of the same taxonomic types tend to distribute in narrow regions in the $ \alpha_0 $--$ P_\mathrm{min} $ plot, suggesting that the distribution in the plot is mostly determined by the compositions rather than the surficial textures for these observed samples. Our data point for 2005 UD is located in the M-type (possibly made of nickel-iron having moderately red spectra, \citealt{1984PhDT.........3T}) concentration and close to the B-type concentration. This similarity can be explained by comparable albedo values (i.e., $ 0.14 \pm 0.04 $ for B-type and $0.13 \pm 0.05 $ for M-type, \citealt{2013Icar..226..723D}). It is, however, unlikely that 2005 UD has an M-type composition because this type of asteroid exhibits slightly red spectra, while 2005 UD exhibits a blue or almost flat spectrum \citep{2006AJ....132.1624J,2007A&A...466.1153K,2020PSJ.....1...15D}. 
Therefore, among asteroids with blue -- flat spectra (indicated as bluer symbols in Figure \ref{fig:compasteroids} (a)), B-type is the best counterpart of 2005 UD in the context of the polarimetric analysis (as described in the previous publications about 2005 UD).

Figure \ref{fig:compasteroids} (b) compares $ \alpha_0 $--$ P_\mathrm{min} $ of 2005 UD with those of meteorite samples. These data were obtained by laboratory light scattering experiments at the University of Arizona \citep{1977LPSC....8.1091Z} and the Meudon Observatory \citep{1986MNRAS.218...75G}. Because the classifications of meteorites have been updated since these publications, we show the latest classification types in Table \ref{table:meteo} based on the web-based service provided by the Meteoritical Society\footnote{https://www.lpi.usra.edu/meteor/metbull.php}. While \citet{1977LPSC....8.1091Z} did not describe the experimental accuracy, \citet{1986MNRAS.218...75G} described the accuracies of $P_\mathrm{min}$ and $h$ ($\pm 0.05$ \% and $\pm$0.005 \% deg$^{-1}$, respectively). These accuracies are high enough for the discussion below. We did not plot the data for achondrite samples herein because their $ P_\mathrm{min} $ values ($ \geq -0.5 $\%) were substantially different from that of 2005 UD. From the comparison, two meteorite samples, CK4 (Karoonda) and EH4 (Abee), exhibit $ \alpha_0 $--$ P_\mathrm{min} $ values similar to that of 2005 UD. However, it is unlikely that 2005 UD has a composition similar to EH4 (Abee), which exhibits a red spectrum in the $ B $ and $ V $ bands \citep{2018JQSRT.206..189P} and therefore disagree with the optical colour of 2005 UD. In Figure \ref{fig:compasteroids} (b), four CV and CO chondrite samples (Allende (CV3), Grosnaja (CV3), Felix (CO3) and Ornans (CO3)) have $ P_\mathrm{min} $ values roughly consistent with that of 2005 UD but indicate significantly large $ \alpha_0 $ values ($ \geq 25 \degr $). In \citet{1977LPSC....8.1091Z}, where these meteorite data were given, the authors refrained from any specific interpretation of the large $ \alpha_0 $ values of these meteorites because $ \alpha_0 $ is sensitive to the sample preparation and the existence of submicron grains rather than the types of meteorites. Thus, at this stage of the discussion, we leave open the possibility that 2005 UD has a composition comparable to those of CV and CO as well as CK.


\begin{table*}
\caption{Polarimetric parameters of the meteorite samples}
\label{table:meteo}
\begin{tabular}{lrrrrrrr}
\hline
 Name & Class & $ P_\mathrm{min} $ (\%) & $ \alpha_0 $ (deg) & $h$ (\% deg$^{-1}$) & Filter$^{*1}$ & Location $^{*2}$ & Reference\\
\hline

Mighei & CM2 & -2.00 & 22.5 & 0.320 & \ldots (0.580) &  \ldots & \citet{1986MNRAS.218...75G} \\
Orgueil & CI1 & -2.00 & 21.5 & 0.240 & \ldots (0.580) & \ldots & \citet{1986MNRAS.218...75G} \\
Murchison & CM2 & -1.80 & 23.0 & 0.300 & \ldots (0.580) & \ldots & \citet{1986MNRAS.218...75G} \\
Allende & CV3 & -1.10 & 25.0 & 0.160 & \ldots (0.580) & \ldots & \citet{1986MNRAS.218...75G} \\
Karoonda & CK4 & -1.00 & 18.5 & 0.130 & \ldots (0.580) & \ldots & \citet{1986MNRAS.218...75G} \\
Abee & EH4 & -1.20 & 18.5 & 0.150 & \ldots (0.580) & \ldots & \citet{1986MNRAS.218...75G} \\
Ochansk & H4 & -0.60 & 19.0 & 0.090 & \ldots (0.580) & \ldots & \citet{1986MNRAS.218...75G} \\
Daniel's Kuil & EL6 & -0.35 & 18.0 &  & \ldots (0.580) & \ldots & \citet{1986MNRAS.218...75G} \\
Oubari & LL6 & -0.65 & 22.5 & 0.080 & \ldots (0.580) & \ldots & \citet{1986MNRAS.218...75G} \\
Pultusk & H5 & -0.60 & 22.0 & 0.050 & \ldots (0.580) & \ldots & \citet{1986MNRAS.218...75G} \\
Girgenti & L6 & -0.35 & 18.0 & 0.040 & \ldots (0.580) & \ldots & \citet{1986MNRAS.218...75G} \\
 &  &  &  &  &  &  &  \\
Mighei & CM2 & -2.00 & 22.8 & 0.327 & O (0.585) & Meudon & \citet{1977LPSC....8.1091Z} \\
Murchison & CM2 & -1.90 & 22.8 & 0.317 & O (0.585) & Meudon & \citet{1977LPSC....8.1091Z} \\
Orgueil & CI1 & -1.90 & 21.9 & 0.208 & O (0.585) & Meudon & \citet{1977LPSC....8.1091Z} \\
Allende & CV3 & -1.20 & 24.8 & 0.158 & O (0.585) & Meudon & \citet{1977LPSC....8.1091Z} \\
Abee & EH4 & -1.20 & 18.3 & 0.147 & O (0.585) & Meudon & \citet{1977LPSC....8.1091Z} \\
Pultusk & H5 & -0.60 & 22.8 & 0.050 & O (0.585) & Meudon & \citet{1977LPSC....8.1091Z} \\
Pultusk & H5 & -0.60 & 22.8 & 0.057 & G (0.520) & Meudon & \citet{1977LPSC....8.1091Z} \\
Girgenti & L6 & -0.40 & 21.6 & 0.039 & O (0.585) & Meudon & \citet{1977LPSC....8.1091Z} \\
Kapoeta & Howardite & -0.50 & 22.0 & 0.048 & O (0.585) & Meudon & \citet{1977LPSC....8.1091Z} \\
Kapoeta & Howardite & -0.50 & 22.0 & 0.052 & G (0.520) & Meudon & \citet{1977LPSC....8.1091Z} \\
Tatahouine & Diogenite & -0.30 & 26.0 & 0.026 & O (0.585) & Meudon & \citet{1977LPSC....8.1091Z} \\
 &  &  &  &  &  &  &  \\
Nogoya & CM2 & -2.20 & 22.5 & 0.285 & G (0.520) & Arizona & \citet{1977LPSC....8.1091Z} \\
Felix & CO3 & -1.40 & 27.5 & 0.174 & G (0.520) & Arizona & \citet{1977LPSC....8.1091Z} \\
Grosnaja & CV3 & -1.45 & 26.2 & 0.169 & G (0.520) & Arizona & \citet{1977LPSC....8.1091Z} \\
Ornans & CO3.4 & -1.50 & 28.4 & 0.126 & G (0.520) & Arizona & \citet{1977LPSC....8.1091Z} \\
Karoonda & CK4 & -1.15 & 19.1 & 0.180 & G (0.520) & Arizona & \citet{1977LPSC....8.1091Z} \\
Paragould & LL5 & -1.10 & 22.0 & 0.129 & G (0.520) & Arizona & \citet{1977LPSC....8.1091Z} \\
Farmington & L5 &  & 19.2 & 0.115 & G (0.520) & Arizona & \citet{1977LPSC....8.1091Z} \\
Tieschitz & H/L3.6 & 0.65 & 20.6 & 0.098 & G (0.520) & Arizona & \citet{1977LPSC....8.1091Z} \\
Olivenza & LL5 & -0.35 & 17.4 & 0.057 & G (0.520) & Arizona & \citet{1977LPSC....8.1091Z} \\
Colby (Wisconsin) & L6 & -0.40 & 18.3 & 0.054 & G (0.520) & Arizona & \citet{1977LPSC....8.1091Z} \\
Pavlovka & Howardite & -0.50 & 18.7 & 0.052 & G (0.520) & Arizona & \citet{1977LPSC....8.1091Z} \\
Nobleborough & Eucrite-pmict & -0.50 & 20.2 & 0.050 & G (0.520) & Arizona & \citet{1977LPSC....8.1091Z} \\
Chassigny & Martian (chassignite) & -0.20 & 17.2 & 0.034 & G (0.520) & Arizona & \citet{1977LPSC....8.1091Z} \\
Norton County & Aubrite & -0.28 & 21.1 & 0.026 & G (0.520) & Arizona & \citet{1977LPSC....8.1091Z} \\

\hline
\multicolumn{8}{l}{Accuracies of data in \citet{1986MNRAS.218...75G} are 0.05 \% for $P_\mathrm{min}$ and 0.005 \% deg$^{-1}$ for $h$.}\\
\multicolumn{8}{l}{The other errors are not written in these reference papers.}\\
\multicolumn{8}{l}{$^{*1}$ Filter name (the central wavelength in $\mu$m). '\ldots' denotes no information in the reference.}\\
\multicolumn{8}{l}{$^{*2}$ Location of the laboratory, either Meudon Observatory or University of Arizona.}
\end{tabular}
\end{table*}

\subsection{\texorpdfstring{$h$}{h}--\texorpdfstring{$ P_\mathrm{min} $}{Pmin} relation}
\label{subsection:hP}
\begin{figure}
\includegraphics[width=\columnwidth]{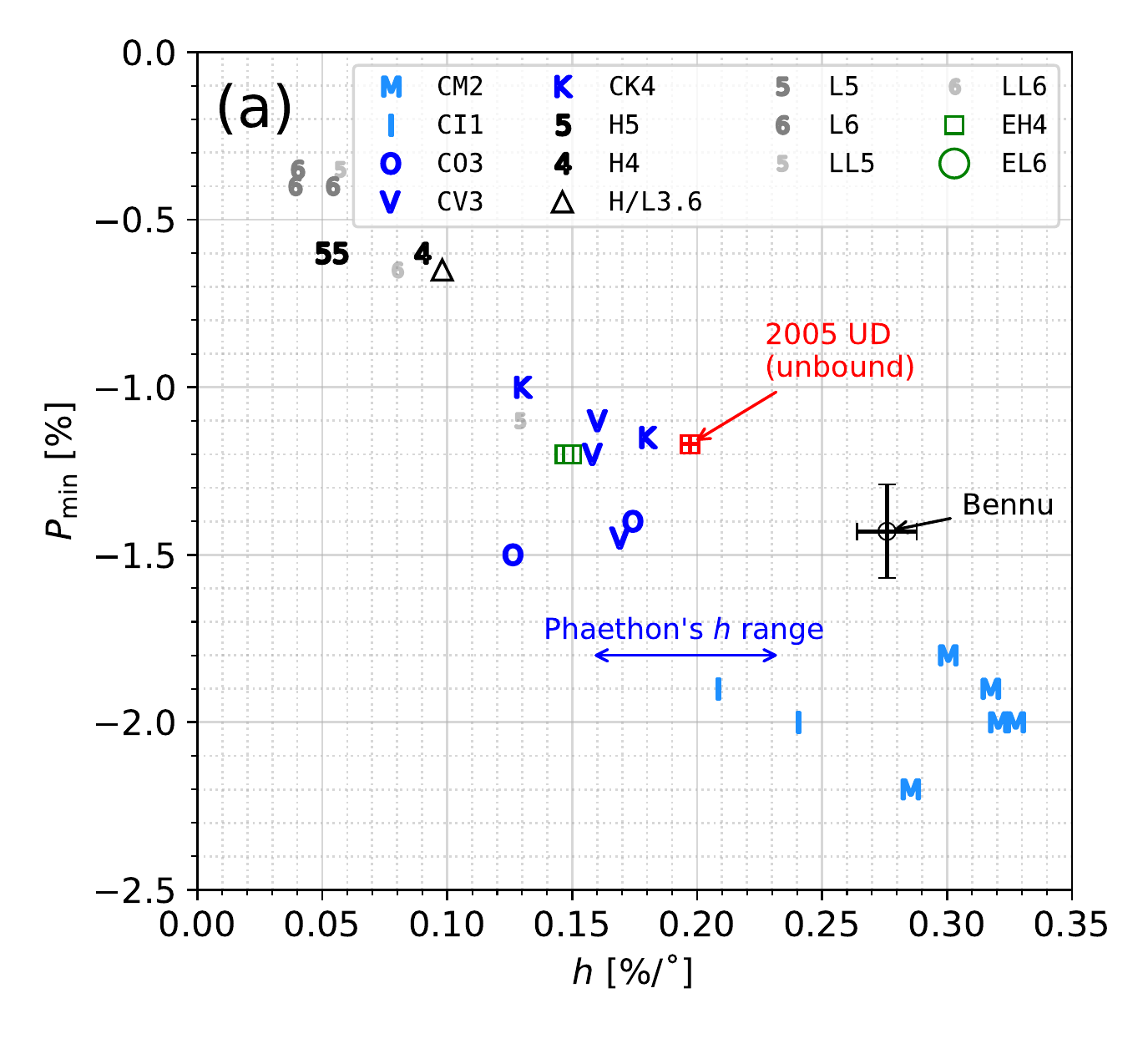}\\
\caption{Comparison of $ h \mathrm{-} P_\mathrm{min} $ between 2005 UD and meteoritic samples. The slope of Phaethon (with an unknown $ P_\mathrm{min} $) is indicated by the arrow based on the result in \citet{2018ApJ...864L..33S}. The B-type NEA, Bennu, is also shown in these plots. The 2005 UD result is derived from all data for the unbound case with MC results in Table \ref{table:polfits}. }
\label{fig:compasteroids2}
\end{figure}

Next, we compared the $ h \mathrm{-} P_\mathrm{min} $ relation of 2005 UD with those of meteoritic samples (Figure \ref{fig:compasteroids2}), where $h$ denotes the polarimetric slope parameter (see Eq. (\ref{eq:LM})). In the plot, we included two B-type asteroids, Phaethon \citep{2018ApJ...864L..33S} and Bennu \citep{2018MNRAS.481L..49C}, for comparison, although Phaethon's $ P_\mathrm{min} $ value has not been determined to date. As we mentioned in Section \ref{subsec:polresult}, $ h $ is a good proxy for geometric albedo, showing that samples with high albedos are distributed leftward, while those with lower albedos are distributed rightward. 
From the comparison between these B-type asteroids and meteoritic samples, we found that 2005 UD and Phaethon are located near the concentration of meteoritic samples of petrographic types 3--4 (CK, CO, and CV, anhydrous). In contrast, we note that Bennu is close to the concentration of samples of anhydrous CK4, CV3, and CO3 chondrites. \citet{2019Natur.568...55L} and \citet{2019NatAs...3..332H} reported that Bennu is linked to CM chondrites, which is consistent with the $ h \mathrm{-} P_\mathrm{min} $ of Bennu in Figure \ref{fig:compasteroids2}. 
Therefore, we expect that the $ h \mathrm{-} P_\mathrm{min} $ relation of C-complex asteroids (including B-type asteroids) would provide a useful measure of aqueous alternation in future research.

\subsection{\texorpdfstring{albedo--$ P_\mathrm{max} $}{Pmax} relation}

\begin{figure*}
  \centering
   \includegraphics[width=1.5\columnwidth]{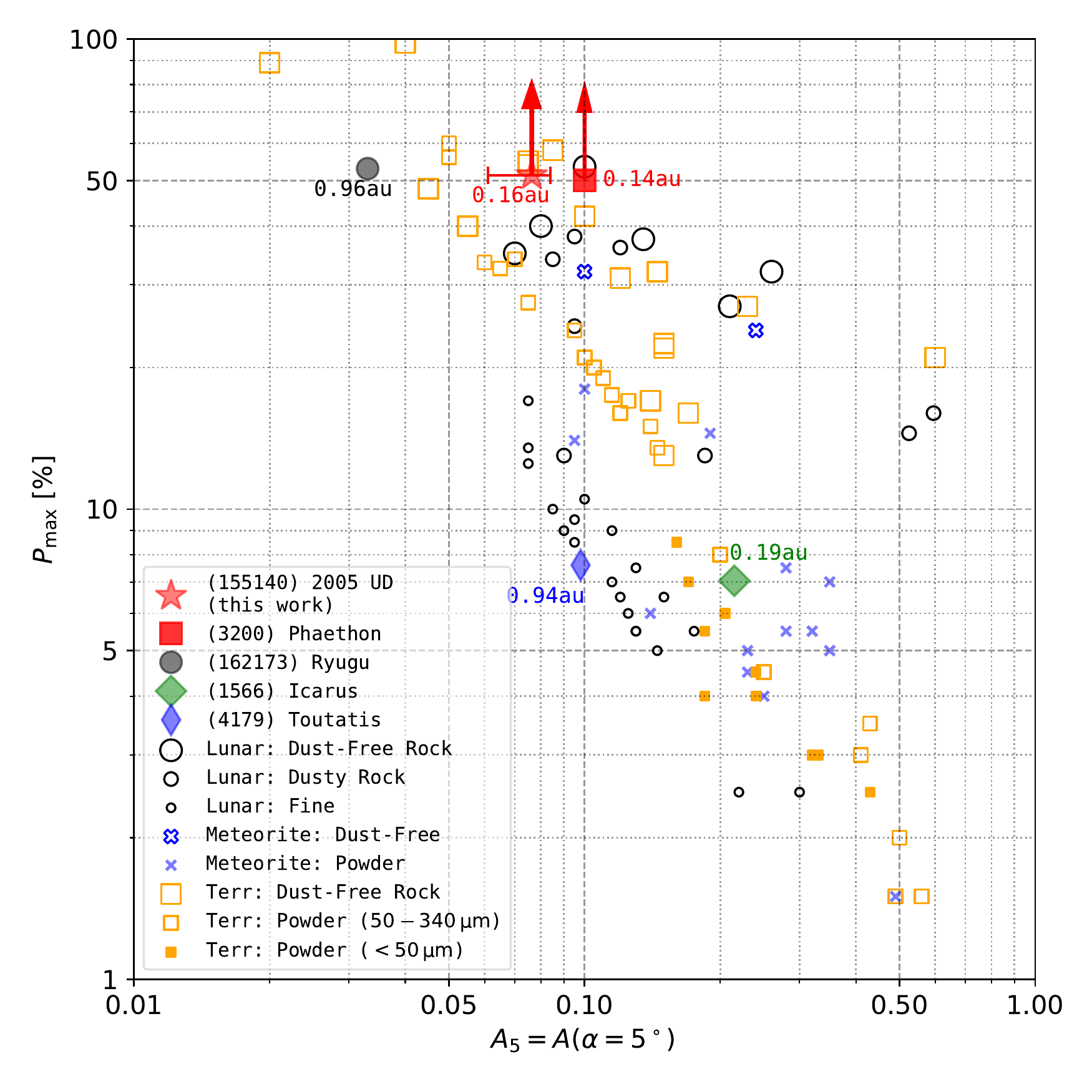}
  \caption{Plot of the albedo at phase angle $ 5\degr $ ($ A_5 $)  and the maximum polarization ($ P_\mathrm{max} $) for laboratory samples (tables in \citealt{1986MNRAS.218...75G}). Overplotted are asteroids, including (162173) Ryugu \citep{2021ApJ...911L..24K}, (3200) Phaethon \citep{2018NatCo...9.2486I}, (1566) Icarus \citep{2017AJ....154..180I}, and (4179) Toutatis \citep{2019JKAS...52...71B}.
   The numbers near the three asteroidal objects denote the perihelion distances of each asteroid in au. }
  \label{fig:figpmax-a5}
\end{figure*}

It is known that there is a correlation between albedo and $ P_\mathrm{max} $ (Umov's law, \citet{1905Umov}) . The albedo--$ P_\mathrm{max} $ relation also depends on particle size and porosity \citealt{1986MNRAS.218...75G,1999Icar..142..281W}. The particle size distribution of the lunar regolith has been investigated by measuring albedo and $ P_\mathrm{max} $ \citep{1999Icar..140..313D,2018ApJ...869...67J}. Although the $ P_\mathrm{max} $ measurements are important for estimating the particle size and porosity, it is not easy to derive the $ P_\mathrm{max} $ of asteroids because of the low visibility from ground-based observatories. Only NEAs provide opportunities to be observed at large phase angles. Figure \ref{fig:figpmax-a5} compares the albedo--$ P_\mathrm{max}$ relation between laboratory samples and asteroids, where the albedo is defined at the phase angle $\alpha$=5\degr. Note that we indicate lower limits of $P_\mathrm{max}$ for Phaethon and 2005 UD. We only consider asteroids with known albedo values observed at phase angle $ \gtrsim 100\degr $ because most solar system airless bodies (such as the Moon, Mercury, cometary dust, and the asteroid 4179 Toutatis) exhibit polarization maxima around $ \alpha \sim 100\degr $, so that these polarimetric data provide more reliable estimates of $ P_\mathrm{max} $ values. Phaethon and 2005 UD are likely covered with rock samples based on visual inspection in Figure \ref{fig:figpmax-a5}.

Using a formula in \citet{1992Icar...99..468S}, we substituted the albedo at $\alpha$=5\degr ($A_5$) and $P_\mathrm{max}$ of 2005 UD and obtained the lower limit of particle size of 280 \micron. The lower limit is close to Phaethon (360 \micron, \citealt{2018NatCo...9.2486I}) and considerably larger than Toutatis ($<$50 \micron, \citealt{2019JKAS...52...71B}). However, it should be noted that these sizes are estimated using a formula based on the lunar regolith experiment. For carbonaceous asteroids such as Phaethon and 2005 UD, this size estimation method may not be applicable because the different composition and microscopic/sub-microscopic structure would result in different polarization phase curves. Therefore, we compare our observational results with the polarimetric measurement of CV and CO carbonaceous chondrites \citep{2011JQSRT.112.1881H,2019MNRAS.484.2198F}. We are unable to find the experimental data for CK chondrites. Comparing $P_\mathrm{max}$ values between 2005 UD, Phaethon, and these anhydrous samples would make sense based on the low phase angle polarimetric properties (see Sections \ref{subsection:aP} and \ref{subsection:hP}). \citet{2011JQSRT.112.1881H} measured ground and sieved samples whose maximum particle sizes were controlled by their device, while minimum sizes were not, and found that $P_\mathrm{max}$ depends on the maximum particle size but does not exceed 29\% (for Allende, CV3 chondrite with the particle size $< 500 \,\micron$). \citet{2019MNRAS.484.2198F} conducted a similar laboratory experiment for CV (Allende and DaG521) and CO3 (FRO99040 and FRO95002) chondrites with effective radius of $ 3.58  \mathrm{-} 8.69 \,\micron $, and derived $P_\mathrm{max} =10.0\mathrm{-}12.6 $\%. None of these ground and sieved carbonaceous chondrite samples showed $P_\mathrm{max}$ values as large as 2005 UD and Phaethon. 

Why are $P_\mathrm{max}$ values of 2005 UD and Phaethon significantly larger than those of these anhydrous carbonaceous chondrites? First, different albedo values could be a possible reason. However, these meteoritic samples indicate albedo values which are almost consistent with 2005 UD and Phaethon. The polarimetric slope $h$ (the proxy of albedo) of the largest ($<500 \,\micron $) CV3 samples ($h=0.18\pm0.02$) is as large as those of 2005 UD ($h=0.192\pm0.006$, this work) and Phaethon \citep[$h = 0.174 \pm 0.053$,][]{2018ApJ...864L..33S}. This fact suggests that the albedos of these asteroids and carbonaceous chondrites are similar to one another. Another possibility is the difference in porosity. According to \citet{2002P&SS...50..895H}, it has been experimentally demonstrated that single-scattering becomes dominant in the case of materials with high porosity; therefore, $ P_\mathrm{max} $ increases. However, in this case, $ \alpha_\mathrm{max}$ becomes $\sim$90\degr, which is not consistent with the polarization phase curve of Phaethon and 2005 UD that keep increasing even when $\alpha > 100\degr$. From a numerical simulation, \citet{2006JQSRT.100..220L} found that the presence of micron-sized grains in fluffy aggregates decreases $P_\mathrm{max}$. Moreover, \citet{2018ApJS..235...19E} conducted a laboratory experiment for a lunar regolith simulant and found that $P_\mathrm{max}$ increased after removing particles with radius smaller than 1 \micron. We thus suspect that such small grains would have been removed from these NSAs to explain the large difference in $P_\mathrm{max}$ between these experiments and observations. 

Therefore, the most plausible explanation for the very large $P_\mathrm{max}$ values with large $\alpha_\mathrm{max}$ of these NSAs is the existence of large grains and the paucity of small micron-sized grains. The ejection of micron-sized grains can explain the lack of small grains via electrostatic lofting \citep{2016JGRE..121.2150Z} or thermal radiation pressure \citep{Bach2021}. However, these mechanisms are not sufficient to explain the dominance of large grains. We conjecture that sintering is a more probable mechanism for modifying the surfaces of these NSAs. The surface temperature of these asteroids reaches around 1000 K (\citealt{2020arXiv201010633M}), which is high enough for the sintering of chondrites (beyond $600 \mathrm{-} 700$ K, \citealt{1984E&PSL..68...34Y,2010JGRE..115.8001G}). 

\vspace{5mm}
By summarizing the polarization properties, the following evolutionary history of these NSAs can be inferred. After these asteroids were injected in the near-Sun orbits, the carbonaceous materials (with either hydrous or anhydrous silicates) would have experienced a high temperature of 900--1000 K around their perihelia. If the original ingredients contained hydrated silicates, they would have fully transformed to anhydrous silicates under such high temperatures \citep[i.e., $>900$ K,][]{1996M&PS...31..321H}.
This is the reason why our polarimetry of NSAs indicates the similarity to anhydrous meteoritic samples that have experienced significant heating.
Small micron--submicron particles which were generated by impacts and thermal stress fill in the gaps between larger particles to produce even larger particles and reduce the porosity by sintering \citep{1984E&PSL..68...34Y}. As the porosity within particles or in the regolith layers decreases and the abundance of micron and submicron-sized grains decreases, the contact areas would enlarge to produce large grains, and, eventually, the multiple scattering is suppressed. This is a possible reason why these NSAs have very large $P_\mathrm{max}$ values and large $\alpha_\mathrm{max}$ despite the fact that their albedos are not as small as hydrated asteroids.

\section{Summary}
We conducted photometric and polarimetric observations of 2005 UD during the 2018 observation opportunity. Our findings are as follows:
\begin{enumerate}
\item the polarization phase curve is similar to that of Phaethon observed in 2016 for a wide range of the observed solar phase angles ($ \alpha = 20 \mathrm{-} 105\degr $) but different from hydrous asteroids (101955) Bennu and (162173) Ryugu,
\item at low phase angles, the polarimetric property of 2005 UD is consistent with anhydrous carbonaceous chondrites,
\item the geometric albedo is in the range from $\pVMin \mathrm{-} \pVMax $, which is consistent with that of Phaethon but significantly larger than those of (101955) Bennu and (162173) Ryugu,
\item the mean absolute magnitude, synodic rotational period, and mean effective diameter are $ H_\mathrm{V} = 17.54 \pm 0.02 $, $ T_\mathrm{rot} = 5.2388 \pm 0.0022 $ hours (assuming that one rotation creates two peaks and two troughs), and $ D_\mathrm{eff} = 1.32 \pm 0.06 \,\mathrm{km} $,
\item at large phase angles, 2005 UD show a polarization degree which is significantly larger than the value of $< 500 \,\micron $ for anhydrous carbonaceous chondrite samples,
\end{enumerate}

We conjecture that the discrepancy in the polarization phase curves at large phase angles can be explained by a dominance of large particles and a paucity of small grains, probably caused by the sintering under the strong solar radiation field.

\section*{Acknowledgments}

This work was supported by the Seoul National University Research Grant in 2018. The Pirka telescope is operated by Graduate School of Science, Hokkaido
University, and is partially supported by the Optical \& Near-Infrared
Astronomy Inter-University Cooperation Program, MEXT, of Japan. MIm and JGS acknowledge the support from the Korea Astronomy and Space Science Institute under the R\&D project, "Korean Small Telescope Network", supervised by the Ministry of Science and ICT. MIm also acknowledges the support from the Korea Astronomy and Space Science Institute grant under the R\&D program (Project No. 2020-1-600-05) supervised by the Ministry of Science and Technology and ICT (MSIT), and the National Research Foundation of Korea (NRF) grant, No. 2020R1A2C3011091, funded by MSIT. We thank Dr. M. Kokubo (Tohoku University) for sharing the MSI polarimetric calibration data taken in 2018 March. Partially based on observations made with the Nordic Optical Telescope, owned in collaboration by the University of Turku and Aarhus University, and operated jointly by Aarhus University, the University of Turku, and the University of Oslo, representing Denmark, Finland, and Norway, the University of Iceland and Stockholm University at the Observatorio del Roque de los Muchachos, La Palma, Spain, of the Instituto de Astrofisica de Canarias. The data presented here were obtained in part with ALFOSC, which is provided by the Instituto de Astrofisica de Andalucia (IAA) under a joint agreement with the University of Copenhagen and NOT. The asteroid's database for $ \alpha_0 $--$ P_\mathrm{min} $ values is provided by courtesy of Dr. Irina Belskaya (V. N. Karazin Kharkiv National University). We appreciate personal discussions with J. Hanu\v{s} about the size and albedo and Edith Hadamcik about polarimetric measurements of meteoritic samples . S.H. was supported by the Hypervelocity Impact Facility (former name: the Space Plasma Laboratory), ISAS, JAXA. During the polarimetric observation period, MIs was supported by the staff members at Nayoro Observatory, Mr. Y. Yamada, Mr. Y. Murakami, Mr. F. Watanabe, Mr. R. Nagayoshi, and Ms. Y. Kato.

\section*{Data Availability}\label{dataave}
We provide the original observation data, the source codes, scripts, and result files, as well as originally developed packages used for this work. The observational data taken at the Nayoro Observatory are available in the Zenodo Repository\footnote{https://zenodo.org/XXXXXXXXX}. Regarding the observational data taken at the Nordic Optical Telescope and the SNU Astronomical Observatory, please contact Maxime Devog{\`e}le (mdevogele@lowell.edu) and Masateru Ishiguro (ishiguro@snu.ac.kr), respectively. The other materials are available via the GitHub service\footnote{https://github.com/ysBach/IshiguroM\_etal\_155140\_2005UD}. The contents are shown below.


\begin{itemize}
\item \texttt{MSI\_NOT}: Data analysis pipeline for the polarimetric data of NO and NOT. The star subtraction code is included.
\item \texttt{polarimetry}: Files related to polarimetric curve fitting (Sect. \ref{subsec:polresult} and Appendix \ref{app: fitting}), plots using polarimetric parameters (Sect \ref{subsection:aP} and \ref{subsection:hP} and Fig. \ref{fig:fig1}, \ref{fig:compasteroids}, and \ref{fig:compasteroids2}).
\item \texttt{photometry}: Files related to photometric data reduction (Sect. \ref{subsec:photoresult}) and light curve analysis (Fig. \ref{fig:lightcurve}).
\item \texttt{data}: The data files that we used in \texttt{polarimetry}.
, except for Fig \ref{fig:compasteroids} (a).
\end{itemize}



\bibliographystyle{mnras}
\bibliography{reference} 

\newcommand{\noop}[1]{}
\begin{thebibliography}{}
\makeatletter
\relax
\def\mn@urlcharsother{\let\do\@makeother \do\$\do\&\do\#\do\^\do\_\do\%\do\~}
\def\mn@doi{\begingroup\mn@urlcharsother \@ifnextchar [ {\mn@doi@}
  {\mn@doi@[]}}
\def\mn@doi@[#1]#2{\def\@tempa{#1}\ifx\@tempa\@empty \href
  {http://dx.doi.org/#2} {doi:#2}\else \href {http://dx.doi.org/#2} {#1}\fi
  \endgroup}
\def\mn@eprint#1#2{\mn@eprint@#1:#2::\@nil}
\def\mn@eprint@arXiv#1{\href {http://arxiv.org/abs/#1} {{\tt arXiv:#1}}}
\def\mn@eprint@dblp#1{\href {http://dblp.uni-trier.de/rec/bibtex/#1.xml}
  {dblp:#1}}
\def\mn@eprint@#1:#2:#3:#4\@nil{\def\@tempa {#1}\def\@tempb {#2}\def\@tempc
  {#3}\ifx \@tempc \@empty \let \@tempc \@tempb \let \@tempb \@tempa \fi \ifx
  \@tempb \@empty \def\@tempb {arXiv}\fi \@ifundefined
  {mn@eprint@\@tempb}{\@tempb:\@tempc}{\expandafter \expandafter \csname
  mn@eprint@\@tempb\endcsname \expandafter{\@tempc}}}

\bibitem[\protect\citeauthoryear{{Ansdell}, {Meech}, {Hainaut}, {Buie},
  {Kaluna}, {Bauer}  \& {Dundon}}{{Ansdell} et~al.}{2014}]{2014ApJ...793...50A}
{Ansdell} M.,  {Meech} K.~J.,  {Hainaut} O.,  {Buie} M.~W.,  {Kaluna} H.,
  {Bauer} J.,   {Dundon} L.,  2014, \mn@doi [\apj]
  {10.1088/0004-637X/793/1/50}, \href
  {https://ui.adsabs.harvard.edu/abs/2014ApJ...793...50A} {793, 50}

\bibitem[\protect\citeauthoryear{{Arai} et~al.,}{{Arai}
  et~al.}{2018}]{2018LPI....49.2570A}
{Arai} T.,  et~al., 2018, in Lunar and Planet. Sci. Conf.. pp No. 2083, id.
  2570

\bibitem[\protect\citeauthoryear{{Bach} \& {Ishiguro}}{{Bach} \&
  {Ishiguro}}{2021}]{Bach2021}
{Bach} Y.~P.,  {Ishiguro} M.,  2021, \aap , in press

\bibitem[\protect\citeauthoryear{{Bach} et~al.,}{{Bach}
  et~al.}{2019}]{2019JKAS...52...71B}
{Bach} Y.~P.,  et~al., 2019, \mn@doi [J. Korean Astron. Soc.]
  {10.5303/JKAS.2019.52.3.71}, \href
  {https://ui.adsabs.harvard.edu/abs/2019JKAS...52...71B} {52, 71}

\bibitem[\protect\citeauthoryear{{Belskaya} et~al.,}{{Belskaya}
  et~al.}{2017}]{2017Icar..284...30B}
{Belskaya} I.~N.,  et~al., 2017, \mn@doi [\icarus]
  {10.1016/j.icarus.2016.11.003}, \href
  {https://ui.adsabs.harvard.edu/abs/2017Icar..284...30B} {284, 30}

\bibitem[\protect\citeauthoryear{{Binzel}, {Rivkin}, {Stuart}, {Harris}, {Bus}
  \& {Burbine}}{{Binzel} et~al.}{2004}]{2004Icar..170..259B}
{Binzel} R.~P.,  {Rivkin} A.~S.,  {Stuart} J.~S.,  {Harris} A.~W.,  {Bus}
  S.~J.,   {Burbine} T.~H.,  2004, \mn@doi [\icarus]
  {10.1016/j.icarus.2004.04.004}, \href
  {https://ui.adsabs.harvard.edu/abs/2004Icar..170..259B} {170, 259}

\bibitem[\protect\citeauthoryear{{Borisov} et~al.,}{{Borisov}
  et~al.}{2018}]{2018MNRAS.480L.131B}
{Borisov} G.,  et~al., 2018, \mn@doi [\mnras] {10.1093/mnrasl/sly140}, \href
  {https://ui.adsabs.harvard.edu/abs/2018MNRAS.480L.131B} {480, L131}

\bibitem[\protect\citeauthoryear{{Buratti}, {Hicks}, {Soderblom}, {Britt},
  {Oberst}  \& {Hillier}}{{Buratti} et~al.}{2004}]{2004Icar..167...16B}
{Buratti} B.~J.,  {Hicks} M.~D.,  {Soderblom} L.~A.,  {Britt} D.,  {Oberst} J.,
    {Hillier} J.~K.,  2004, \mn@doi [\icarus] {10.1016/j.icarus.2003.05.002},
  \href {https://ui.adsabs.harvard.edu/abs/2004Icar..167...16B} {167, 16}

\bibitem[\protect\citeauthoryear{{Cellino}, {Bagnulo}, {Gil-Hutton}, {Tanga},
  {Ca{\~n}ada-Assandri}  \& {Tedesco}}{{Cellino}
  et~al.}{2015}]{2015MNRAS.451.3473C}
{Cellino} A.,  {Bagnulo} S.,  {Gil-Hutton} R.,  {Tanga} P.,
  {Ca{\~n}ada-Assandri} M.,   {Tedesco} E.~F.,  2015, \mn@doi [\mnras]
  {10.1093/mnras/stv1188}, \href
  {https://ui.adsabs.harvard.edu/abs/2015MNRAS.451.3473C} {451, 3473}

\bibitem[\protect\citeauthoryear{{Cellino}, {Bagnulo}, {Belskaya}  \&
  {Christou}}{{Cellino} et~al.}{2018}]{2018MNRAS.481L..49C}
{Cellino} A.,  {Bagnulo} S.,  {Belskaya} I.~N.,   {Christou} A.~A.,  2018,
  \mn@doi [\mnras] {10.1093/mnrasl/sly156}, \href
  {https://ui.adsabs.harvard.edu/abs/2018MNRAS.481L..49C} {481, L49}

\bibitem[\protect\citeauthoryear{{Ciarniello} et~al.,}{{Ciarniello}
  et~al.}{2015}]{2015A&A...583A..31C}
{Ciarniello} M.,  et~al., 2015, \mn@doi [\aap] {10.1051/0004-6361/201526307},
  \href {https://ui.adsabs.harvard.edu/abs/2015A&A...583A..31C} {583, A31}

\bibitem[\protect\citeauthoryear{{DeMeo} \& {Carry}}{{DeMeo} \&
  {Carry}}{2013}]{2013Icar..226..723D}
{DeMeo} F.~E.,  {Carry} B.,  2013, \mn@doi [\icarus]
  {10.1016/j.icarus.2013.06.027}, \href
  {https://ui.adsabs.harvard.edu/abs/2013Icar..226..723D} {226, 723}

\bibitem[\protect\citeauthoryear{{DeMeo}, {Binzel}, {Slivan}  \& {Bus}}{{DeMeo}
  et~al.}{2009}]{2009Icar..202..160D}
{DeMeo} F.~E.,  {Binzel} R.~P.,  {Slivan} S.~M.,   {Bus} S.~J.,  2009, \mn@doi
  [\icarus] {10.1016/j.icarus.2009.02.005}, \href
  {https://ui.adsabs.harvard.edu/abs/2009Icar..202..160D} {202, 160}

\bibitem[\protect\citeauthoryear{{Devog{\`e}le} et~al.,}{{Devog{\`e}le}
  et~al.}{2018}]{2018MNRAS.479.3498D}
{Devog{\`e}le} M.,  et~al., 2018, \mn@doi [\mnras] {10.1093/mnras/sty1587},
  \href {https://ui.adsabs.harvard.edu/abs/2018MNRAS.479.3498D} {479, 3498}

\bibitem[\protect\citeauthoryear{{Devog{\`e}le} et~al.,}{{Devog{\`e}le}
  et~al.}{2020}]{2020PSJ.....1...15D}
{Devog{\`e}le} M.,  et~al., 2020, \mn@doi [Planet. Sci. J.]
  {10.3847/PSJ/ab8e45}, \href
  {https://ui.adsabs.harvard.edu/abs/2020PSJ.....1...15D} {1, 15}

\bibitem[\protect\citeauthoryear{{Dollfus}}{{Dollfus}}{1999}]{1999Icar..140..313D}
{Dollfus} A.,  1999, \mn@doi [\icarus] {10.1006/icar.1999.6145}, \href
  {https://ui.adsabs.harvard.edu/abs/1999Icar..140..313D} {140, 313}

\bibitem[\protect\citeauthoryear{{Dollfus} \& {Geake}}{{Dollfus} \&
  {Geake}}{1975}]{1975LPSC....6.2749D}
{Dollfus} A.,  {Geake} J.~E.,  1975, Lunar and Planet. Sci. Conf. Proc., \href
  {https://ui.adsabs.harvard.edu/abs/1975LPSC....6.2749D} {3, 2749}

\bibitem[\protect\citeauthoryear{{Dollfus} \& {Titulaer}}{{Dollfus} \&
  {Titulaer}}{1971}]{1971A&A....12..199D}
{Dollfus} A.,  {Titulaer} C.,  1971, \aap, \href
  {https://ui.adsabs.harvard.edu/abs/1971A&A....12..199D} {12, 199}

\bibitem[\protect\citeauthoryear{{Escobar-Cerezo} et~al.,}{{Escobar-Cerezo}
  et~al.}{2018}]{2018ApJS..235...19E}
{Escobar-Cerezo} J.,  et~al., 2018, \mn@doi [\apjs] {10.3847/1538-4365/aaa6cc},
  \href {https://ui.adsabs.harvard.edu/abs/2018ApJS..235...19E} {235, 19}

\bibitem[\protect\citeauthoryear{{Fern{\'a}ndez} et~al.,}{{Fern{\'a}ndez}
  et~al.}{2013}]{2013Icar..226.1138F}
{Fern{\'a}ndez} Y.~R.,  et~al., 2013, \mn@doi [\icarus]
  {10.1016/j.icarus.2013.07.021}, \href
  {https://ui.adsabs.harvard.edu/abs/2013Icar..226.1138F} {226, 1138}

\bibitem[\protect\citeauthoryear{{Flewelling} et~al.,}{{Flewelling}
  et~al.}{2020}]{2020ApJS..251....7F}
{Flewelling} H.~A.,  et~al., 2020, \mn@doi [\apjs] {10.3847/1538-4365/abb82d},
  \href {https://ui.adsabs.harvard.edu/abs/2020ApJS..251....7F} {251, 7}

\bibitem[\protect\citeauthoryear{{Fraser} et~al.,}{{Fraser}
  et~al.}{2016}]{2016AJ....151..158F}
{Fraser} W.,  et~al., 2016, \mn@doi [\aj] {10.3847/0004-6256/151/6/158}, \href
  {https://ui.adsabs.harvard.edu/abs/2016AJ....151..158F} {151, 158}

\bibitem[\protect\citeauthoryear{{Frattin} et~al.,}{{Frattin}
  et~al.}{2019}]{2019MNRAS.484.2198F}
{Frattin} E.,  et~al., 2019, \mn@doi [\mnras] {10.1093/mnras/stz129}, \href
  {https://ui.adsabs.harvard.edu/abs/2019MNRAS.484.2198F} {484, 2198}

\bibitem[\protect\citeauthoryear{{Gaia Collaboration} et~al.,}{{Gaia
  Collaboration} et~al.}{2018}]{2018A&A...616A...1G}
{Gaia Collaboration} et~al., 2018, \mn@doi [\aap]
  {10.1051/0004-6361/201833051}, \href
  {https://ui.adsabs.harvard.edu/abs/2018A&A...616A...1G} {616, A1}

\bibitem[\protect\citeauthoryear{{Geake} \& {Dollfus}}{{Geake} \&
  {Dollfus}}{1986}]{1986MNRAS.218...75G}
{Geake} J.~E.,  {Dollfus} A.,  1986, \mn@doi [\mnras] {10.1093/mnras/218.1.75},
  \href {https://ui.adsabs.harvard.edu/abs/1986MNRAS.218...75G} {218, 75}

\bibitem[\protect\citeauthoryear{{Geake} \& {Geake}}{{Geake} \&
  {Geake}}{1990}]{1990MNRAS.245...46G}
{Geake} J.~E.,  {Geake} M.,  1990, \mnras, \href
  {https://ui.adsabs.harvard.edu/abs/1990MNRAS.245...46G} {245, 46}

\bibitem[\protect\citeauthoryear{{Granvik} et~al.,}{{Granvik}
  et~al.}{2016}]{2016Natur.530..303G}
{Granvik} M.,  et~al., 2016, \mn@doi [\nat] {10.1038/nature16934}, \href
  {https://ui.adsabs.harvard.edu/abs/2016Natur.530..303G} {530, 303}

\bibitem[\protect\citeauthoryear{{Gupta} \& {Sahijpal}}{{Gupta} \&
  {Sahijpal}}{2010}]{2010JGRE..115.8001G}
{Gupta} G.,  {Sahijpal} S.,  2010, \mn@doi [\jgr] {10.1029/2009JE003525}, \href
  {https://ui.adsabs.harvard.edu/abs/2010JGRE..115.8001G} {115, E08001}

\bibitem[\protect\citeauthoryear{{Hadamcik}, {Renard}, {Levasseur-Regourd}  \&
  {Worms}}{{Hadamcik} et~al.}{2002}]{2002P&SS...50..895H}
{Hadamcik} E.,  {Renard} J.~B.,  {Levasseur-Regourd} A.~C.,   {Worms} J.~C.,
  2002, \mn@doi [\planss] {10.1016/S0032-0633(02)00065-X}, \href
  {https://ui.adsabs.harvard.edu/abs/2002P&SS...50..895H} {50, 895}

\bibitem[\protect\citeauthoryear{{Hadamcik}, {Levasseur-Regourd}, {Renard},
  {Lasue}  \& {Sen}}{{Hadamcik} et~al.}{2011}]{2011JQSRT.112.1881H}
{Hadamcik} E.,  {Levasseur-Regourd} A.~C.,  {Renard} J.~B.,  {Lasue} J.,
  {Sen} A.~K.,  2011, \mn@doi [\jqsrt] {10.1016/j.jqsrt.2011.01.035}, \href
  {https://ui.adsabs.harvard.edu/abs/2011JQSRT.112.1881H} {112, 1881}

\bibitem[\protect\citeauthoryear{{Hamilton} et~al.,}{{Hamilton}
  et~al.}{2019}]{2019NatAs...3..332H}
{Hamilton} V.~E.,  et~al., 2019, \mn@doi [Nature Astron.]
  {10.1038/s41550-019-0722-2}, \href
  {https://ui.adsabs.harvard.edu/abs/2019NatAs...3..332H} {3, 332}

\bibitem[\protect\citeauthoryear{{Hanu{\v s}} et~al.,}{{Hanu{\v s}}
  et~al.}{2016}]{2016A&A...592A..34H}
{Hanu{\v s}} J.,  et~al., 2016, \mn@doi [\aap] {10.1051/0004-6361/201628666},
  \href {https://ui.adsabs.harvard.edu/abs/2016A%26A...592A..34H} {592, A34}

\bibitem[\protect\citeauthoryear{{Hanu{\v s}} et~al.,}{{Hanu{\v s}}
  et~al.}{2018}]{2018A&A...620L...8H}
{Hanu{\v s}} J.,  et~al., 2018, \mn@doi [\aap] {10.1051/0004-6361/201834228},
  \href {https://ui.adsabs.harvard.edu/abs/2018A%26A...620L...8H} {620, L8}

\bibitem[\protect\citeauthoryear{{Hiroi}, {Zolensky}, {Pieters}  \&
  {Lipschutz}}{{Hiroi} et~al.}{1996}]{1996M&PS...31..321H}
{Hiroi} T.,  {Zolensky} M.~E.,  {Pieters} C.~M.,   {Lipschutz} M.~E.,  1996,
  \mn@doi [Meteoritics Planet. Sci.] {10.1111/j.1945-5100.1996.tb02068.x},
  \href {https://ui.adsabs.harvard.edu/abs/1996M&PS...31..321H} {31, 321}

\bibitem[\protect\citeauthoryear{{Holmberg}, {Flynn}  \&
  {Portinari}}{{Holmberg} et~al.}{2006}]{2006MNRAS.367..449H}
{Holmberg} J.,  {Flynn} C.,   {Portinari} L.,  2006, \mn@doi [\mnras]
  {10.1111/j.1365-2966.2005.09832.x}, \href
  {https://ui.adsabs.harvard.edu/abs/2006MNRAS.367..449H} {367, 449}

\bibitem[\protect\citeauthoryear{{Hsieh} \& {Jewitt}}{{Hsieh} \&
  {Jewitt}}{2005}]{2005ApJ...624.1093H}
{Hsieh} H.~H.,  {Jewitt} D.,  2005, \mn@doi [\apj] {10.1086/429250}, \href
  {https://ui.adsabs.harvard.edu/abs/2005ApJ...624.1093H} {624, 1093}

\bibitem[\protect\citeauthoryear{{Huang}, {Muinonen}, {Chen}  \&
  {Wang}}{{Huang} et~al.}{2021}]{2021P&SS..19505120H}
{Huang} J.~N.,  {Muinonen} K.,  {Chen} T.,   {Wang} X.~B.,  2021, \mn@doi
  [\planss] {10.1016/j.pss.2020.105120}, \href
  {https://ui.adsabs.harvard.edu/abs/2021P&SS..19505120H} {195, 105120}

\bibitem[\protect\citeauthoryear{{Hui} \& {Li}}{{Hui} \&
  {Li}}{2017}]{2017AJ....153...23H}
{Hui} M.-T.,  {Li} J.,  2017, \mn@doi [\aj] {10.3847/1538-3881/153/1/23}, \href
  {https://ui.adsabs.harvard.edu/abs/2017AJ....153...23H} {153, 23}

\bibitem[\protect\citeauthoryear{{Im} et~al.,}{{Im} et~al.}{2021}]{IM:2021}
{Im} M.,  et~al., 2021, Journal of Korean Astronomical Society, \href
  {https://ui.adsabs.harvard.edu/abs/2021JKAS...54...89I} {54, 89}

\bibitem[\protect\citeauthoryear{{Ishiguro}, {Nakayama}, {Kogachi}, {Mukai},
  {Nakamura}, {Hirata}  \& {Okazaki}}{{Ishiguro}
  et~al.}{1997}]{1997PASJ...49L..31I}
{Ishiguro} M.,  {Nakayama} H.,  {Kogachi} M.,  {Mukai} T.,  {Nakamura} R.,
  {Hirata} R.,   {Okazaki} A.,  1997, \mn@doi [\pasj] {10.1093/pasj/49.5.L31},
  \href {https://ui.adsabs.harvard.edu/abs/1997PASJ...49L..31I} {49, L31}

\bibitem[\protect\citeauthoryear{{Ishiguro} et~al.,}{{Ishiguro}
  et~al.}{2017}]{2017AJ....154..180I}
{Ishiguro} M.,  et~al., 2017, \mn@doi [\aj] {10.3847/1538-3881/aa8b1a}, \href
  {https://ui.adsabs.harvard.edu/abs/2017AJ....154..180I} {154, 180}

\bibitem[\protect\citeauthoryear{{Ito} et~al.,}{{Ito}
  et~al.}{2018}]{2018NatCo...9.2486I}
{Ito} T.,  et~al., 2018, \mn@doi [Nature Communications]
  {10.1038/s41467-018-04727-2}, \href
  {https://ui.adsabs.harvard.edu/abs/2018NatCo...9.2486I} {9, 2486}

\bibitem[\protect\citeauthoryear{{Jeong}, {Choi}, {Kim}, {Kim}, {Shkuratov}  \&
  {Yang}}{{Jeong} et~al.}{2018}]{2018ApJ...869...67J}
{Jeong} M.,  {Choi} Y.-J.,  {Kim} S.~S.,  {Kim} I.-H.,  {Shkuratov} Y.~G.,
  {Yang} H.,  2018, \mn@doi [\apj] {10.3847/1538-4357/aae9ed}, \href
  {https://ui.adsabs.harvard.edu/abs/2018ApJ...869...67J} {869, 67}

\bibitem[\protect\citeauthoryear{{Jewitt}}{{Jewitt}}{2012}]{2012AJ....143...66J}
{Jewitt} D.,  2012, \mn@doi [\aj] {10.1088/0004-6256/143/3/66}, \href
  {https://ui.adsabs.harvard.edu/abs/2012AJ....143...66J} {143, 66}

\bibitem[\protect\citeauthoryear{{Jewitt}}{{Jewitt}}{2013}]{2013AJ....145..133J}
{Jewitt} D.,  2013, \mn@doi [\aj] {10.1088/0004-6256/145/5/133}, \href
  {https://ui.adsabs.harvard.edu/abs/2013AJ....145..133J} {145, 133}

\bibitem[\protect\citeauthoryear{{Jewitt} \& {Hsieh}}{{Jewitt} \&
  {Hsieh}}{2006}]{2006AJ....132.1624J}
{Jewitt} D.,  {Hsieh} H.,  2006, \mn@doi [\aj] {10.1086/507483}, \href
  {https://ui.adsabs.harvard.edu/abs/2006AJ....132.1624J} {132, 1624}

\bibitem[\protect\citeauthoryear{{Kareta}, {Reddy}, {Hergenrother}, {Lauretta},
  {Arai}, {Takir}, {Sanchez}  \& {Hanu{\v s}}}{{Kareta}
  et~al.}{2018}]{2018AJ....156..287K}
{Kareta} T.,  {Reddy} V.,  {Hergenrother} C.,  {Lauretta} D.~S.,  {Arai} T.,
  {Takir} D.,  {Sanchez} J.,   {Hanu{\v s}} J.,  2018, \mn@doi [\aj]
  {10.3847/1538-3881/aaeb8a}, \href
  {https://ui.adsabs.harvard.edu/abs/2018AJ....156..287K} {156, 287}

\bibitem[\protect\citeauthoryear{{Kareta}, {Reddy}, {Pearson}, {Sanchez}  \&
  {Harris}}{{Kareta} et~al.}{2021}]{2021arXiv210901020K}
{Kareta} T.,  {Reddy} V.,  {Pearson} N.,  {Sanchez} J.~A.,   {Harris} W.~M.,
  2021, arXiv e-prints, \href
  {https://ui.adsabs.harvard.edu/abs/2021arXiv210901020K} {p. arXiv:2109.01020}

\bibitem[\protect\citeauthoryear{{Kenknight}, {Rosenberg}  \&
  {Wehner}}{{Kenknight} et~al.}{1967}]{1967JGR....72.3105K}
{Kenknight} C.~E.,  {Rosenberg} D.~L.,   {Wehner} G.~K.,  1967, \mn@doi [\jgr]
  {10.1029/JZ072i012p03105}, \href
  {http://adsabs.harvard.edu/abs/1967JGR....72.3105K} {72, 3105}

\bibitem[\protect\citeauthoryear{{Kim}, {Ishiguro}  \& {Usui}}{{Kim}
  et~al.}{2014}]{2014ApJ...789..151K}
{Kim} Y.,  {Ishiguro} M.,   {Usui} F.,  2014, \mn@doi [\apj]
  {10.1088/0004-637X/789/2/151}, \href
  {https://ui.adsabs.harvard.edu/abs/2014ApJ...789..151K} {789, 151}

\bibitem[\protect\citeauthoryear{{Kim} et~al.,}{{Kim}
  et~al.}{2018}]{2018A&A...619A.123K}
{Kim} M.-J.,  et~al., 2018, \mn@doi [\aap] {10.1051/0004-6361/201833593}, \href
  {https://ui.adsabs.harvard.edu/abs/2018A%26A...619A.123K} {619, A123}

\bibitem[\protect\citeauthoryear{{Kinoshita} et~al.,}{{Kinoshita}
  et~al.}{2007}]{2007A&A...466.1153K}
{Kinoshita} D.,  et~al., 2007, \mn@doi [\aap] {10.1051/0004-6361:20066276},
  \href {https://ui.adsabs.harvard.edu/abs/2007A&A...466.1153K} {466, 1153}

\bibitem[\protect\citeauthoryear{{Kiselev}, {Rosenbush}  \&
  {Jockers}}{{Kiselev} et~al.}{1999}]{1999Icar..140..464K}
{Kiselev} N.~N.,  {Rosenbush} V.~K.,   {Jockers} K.,  1999, \mn@doi [\icarus]
  {10.1006/icar.1999.6139}, \href
  {https://ui.adsabs.harvard.edu/abs/1999Icar..140..464K} {140, 464}

\bibitem[\protect\citeauthoryear{{Krugly} et~al.,}{{Krugly}
  et~al.}{2019}]{2019EPSC...13.1989K}
{Krugly} Y.,  et~al., 2019, in EPSC-DPS Joint Meeting 2019. pp
  EPSC--DPS2019--1989

\bibitem[\protect\citeauthoryear{{Kuroda} et~al.,}{{Kuroda}
  et~al.}{2018}]{2018A&A...611A..31K}
{Kuroda} D.,  et~al., 2018, \mn@doi [\aap] {10.1051/0004-6361/201732086}, \href
  {https://ui.adsabs.harvard.edu/abs/2018A%26A...611A..31K} {611, A31}

\bibitem[\protect\citeauthoryear{{Kuroda} et~al.,}{{Kuroda}
  et~al.}{2021}]{2021ApJ...911L..24K}
{Kuroda} D.,  et~al., 2021, \mn@doi [\apjl] {10.3847/2041-8213/abee25}, \href
  {https://ui.adsabs.harvard.edu/abs/2021ApJ...911L..24K} {911, L24}

\bibitem[\protect\citeauthoryear{{Lang}, {Hogg}, {Mierle}, {Blanton}  \&
  {Roweis}}{{Lang} et~al.}{2010}]{2010AJ....139.1782L}
{Lang} D.,  {Hogg} D.~W.,  {Mierle} K.,  {Blanton} M.,   {Roweis} S.,  2010,
  \mn@doi [\aj] {10.1088/0004-6256/139/5/1782}, \href
  {https://ui.adsabs.harvard.edu/abs/2010AJ....139.1782L} {139, 1782}

\bibitem[\protect\citeauthoryear{{Lasue} \& {Levasseur-Regourd}}{{Lasue} \&
  {Levasseur-Regourd}}{2006}]{2006JQSRT.100..220L}
{Lasue} J.,  {Levasseur-Regourd} A.~C.,  2006, \mn@doi [\jqsrt]
  {10.1016/j.jqsrt.2005.11.040}, \href
  {https://ui.adsabs.harvard.edu/abs/2006JQSRT.100..220L} {100, 220}

\bibitem[\protect\citeauthoryear{{Lauretta} et~al.,}{{Lauretta}
  et~al.}{2019}]{2019Natur.568...55L}
{Lauretta} D.~S.,  et~al., 2019, \mn@doi [\nat] {10.1038/s41586-019-1033-6},
  \href {https://ui.adsabs.harvard.edu/abs/2019Natur.568...55L} {568, 55}

\bibitem[\protect\citeauthoryear{{Li} \& {Jewitt}}{{Li} \&
  {Jewitt}}{2013}]{2013AJ....145..154L}
{Li} J.,  {Jewitt} D.,  2013, \mn@doi [\aj] {10.1088/0004-6256/145/6/154},
  \href {https://ui.adsabs.harvard.edu/abs/2013AJ....145..154L} {145, 154}

\bibitem[\protect\citeauthoryear{{Li}, {A'Hearn}, {Farnham}  \&
  {McFadden}}{{Li} et~al.}{2009}]{2009Icar..204..209L}
{Li} J.-Y.,  {A'Hearn} M.~F.,  {Farnham} T.~L.,   {McFadden} L.~A.,  2009,
  \mn@doi [\icarus] {10.1016/j.icarus.2009.06.002}, \href
  {https://ui.adsabs.harvard.edu/abs/2009Icar..204..209L} {204, 209}

\bibitem[\protect\citeauthoryear{{Li}, {A'Hearn}, {Belton}, {Farnham},
  {Klaasen}, {Sunshine}, {Thomas}  \& {Veverka}}{{Li}
  et~al.}{2013}]{2013Icar..222..467L}
{Li} J.-Y.,  {A'Hearn} M.~F.,  {Belton} M. J.~S.,  {Farnham} T.~L.,  {Klaasen}
  K.~P.,  {Sunshine} J.~M.,  {Thomas} P.~C.,   {Veverka} J.,  2013, \mn@doi
  [\icarus] {10.1016/j.icarus.2012.02.011}, \href
  {https://ui.adsabs.harvard.edu/abs/2013Icar..222..467L} {222, 467}

\bibitem[\protect\citeauthoryear{{Lumme} \& {Muinonen}}{{Lumme} \&
  {Muinonen}}{1993}]{1993LPICo.810..194L}
{Lumme} K.,  {Muinonen} K.~O.,  1993, in IAU Symp. 160: Asteroids, Comets,
  Meteors 1993, in Belgirate, Italy. p.~194

\bibitem[\protect\citeauthoryear{{Lupishko}}{{Lupishko}}{2018}]{2018SoSyR..52...98L}
{Lupishko} D.~F.,  2018, \mn@doi [Solar System Res.]
  {10.1134/S0038094618010069}, \href
  {https://ui.adsabs.harvard.edu/abs/2018SoSyR..52...98L} {52, 98}

\bibitem[\protect\citeauthoryear{{Lupishko}, {Vasilyev}, {Efimov}  \&
  {Shakhovskoj}}{{Lupishko} et~al.}{1995}]{1995Icar..113..200L}
{Lupishko} D.~F.,  {Vasilyev} S.~V.,  {Efimov} J.~S.,   {Shakhovskoj} N.~M.,
  1995, \mn@doi [\icarus] {10.1006/icar.1995.1016}, \href
  {https://ui.adsabs.harvard.edu/abs/1995Icar..113..200L} {113, 200}

\bibitem[\protect\citeauthoryear{{MacLennan}, {Toliou}  \&
  {Granvik}}{{MacLennan} et~al.}{2021}]{2020arXiv201010633M}
{MacLennan} E.~M.,  {Toliou} A.,   {Granvik} M.,  2021, \mn@doi [\icarus]
  {10.1016/j.icarus.2021.114535}, 366, 114535

\bibitem[\protect\citeauthoryear{{Masiero}, {Wright}  \& {Mainzer}}{{Masiero}
  et~al.}{2019}]{2019AJ....158...97M}
{Masiero} J.~R.,  {Wright} E.~L.,   {Mainzer} A.~K.,  2019, \mn@doi [\aj]
  {10.3847/1538-3881/ab31a6}, \href
  {https://ui.adsabs.harvard.edu/abs/2019AJ....158...97M} {158, 97}

\bibitem[\protect\citeauthoryear{{Mukai} et~al.,}{{Mukai}
  et~al.}{1997}]{1997Icar..127..452M}
{Mukai} T.,  et~al., 1997, \mn@doi [\icarus] {10.1006/icar.1997.5700}, \href
  {https://ui.adsabs.harvard.edu/abs/1997Icar..127..452M} {127, 452}

\bibitem[\protect\citeauthoryear{{Ohtsuka}, {Sekiguchi}, {Kinoshita}  \&
  {Watanabe}}{{Ohtsuka} et~al.}{2005}]{2005CBET..283....1O}
{Ohtsuka} K.,  {Sekiguchi} T.,  {Kinoshita} D.,   {Watanabe} J.,  2005, Central
  Bureau Electronic Telegrams, \href
  {https://ui.adsabs.harvard.edu/abs/2005CBET..283....1O} {283, 1}

\bibitem[\protect\citeauthoryear{{Ohtsuka}, {Sekiguchi}, {Kinoshita},
  {Watanabe}, {Ito}, {Arakida}  \& {Kasuga}}{{Ohtsuka}
  et~al.}{2006}]{2006A&A...450L..25O}
{Ohtsuka} K.,  {Sekiguchi} T.,  {Kinoshita} D.,  {Watanabe} J.-I.,  {Ito} T.,
  {Arakida} H.,   {Kasuga} T.,  2006, \mn@doi [\aap]
  {10.1051/0004-6361:200600022}, \href
  {https://ui.adsabs.harvard.edu/abs/2006A%26A...450L..25O} {450, L25}

\bibitem[\protect\citeauthoryear{{Ohtsuka}, {Nakato}, {Nakamura}, {Kinoshita},
  {Ito}, {Yoshikawa}  \& {Hasegawa}}{{Ohtsuka}
  et~al.}{2009}]{2009PASJ...61.1375O}
{Ohtsuka} K.,  {Nakato} A.,  {Nakamura} T.,  {Kinoshita} D.,  {Ito} T.,
  {Yoshikawa} M.,   {Hasegawa} S.,  2009, \mn@doi [\pasj]
  {10.1093/pasj/61.6.1375}, \href
  {https://ui.adsabs.harvard.edu/abs/2009PASJ...61.1375O} {61, 1375}

\bibitem[\protect\citeauthoryear{{Okazaki} et~al.,}{{Okazaki}
  et~al.}{2020}]{2020P&SS..18004774O}
{Okazaki} R.,  et~al., 2020, \mn@doi [\planss] {10.1016/j.pss.2019.104774},
  \href {https://ui.adsabs.harvard.edu/abs/2020P&SS..18004774O} {180, 104774}

\bibitem[\protect\citeauthoryear{{Penttil{\"a}}, {Martikainen}, {Gritsevich}
  \& {Muinonen}}{{Penttil{\"a}} et~al.}{2018}]{2018JQSRT.206..189P}
{Penttil{\"a}} A.,  {Martikainen} J.,  {Gritsevich} M.,   {Muinonen} K.,  2018,
  \mn@doi [\jqsrt] {10.1016/j.jqsrt.2017.11.011}, \href
  {https://ui.adsabs.harvard.edu/abs/2018JQSRT.206..189P} {206, 189}

\bibitem[\protect\citeauthoryear{{Pravec} \& {Harris}}{{Pravec} \&
  {Harris}}{2007}]{2007Icar..190..250P}
{Pravec} P.,  {Harris} A.~W.,  2007, \mn@doi [\icarus]
  {10.1016/j.icarus.2007.02.023}, \href
  {https://ui.adsabs.harvard.edu/abs/2007Icar..190..250P} {190, 250}

\bibitem[\protect\citeauthoryear{Press, Teukolsky, Vetterling  \&
  Flannery}{Press et~al.}{2007}]{press2007numerical}
Press W.~H.,  Teukolsky S.~A.,  Vetterling W.~T.,   Flannery B.~P.,  2007,
  Numerical recipes 3rd edition: The art of scientific computing.
Cambridge university press

\bibitem[\protect\citeauthoryear{{Ryabova}, {Avdyushev}  \&
  {Williams}}{{Ryabova} et~al.}{2019}]{2019MNRAS.485.3378R}
{Ryabova} G.~O.,  {Avdyushev} V.~A.,   {Williams} I.~P.,  2019, \mn@doi
  [\mnras] {10.1093/mnras/stz658}, \href
  {https://ui.adsabs.harvard.edu/abs/2019MNRAS.485.3378R} {485, 3378}

\bibitem[\protect\citeauthoryear{{Salvatier}, {Wiecki{\^a}}  \&
  {Fonnesbeck}}{{Salvatier} et~al.}{2016}]{2016ascl.soft10016S}
{Salvatier} J.,  {Wiecki{\^a}} T.~V.,   {Fonnesbeck} C.,  2016, {PyMC3: Python
  probabilistic programming framework} (\mn@eprint {ascl} {1610.016})

\bibitem[\protect\citeauthoryear{{Shinnaka}, {Kasuga}, {Furusho}, {Boice},
  {Terai}, {Noda}, {Namiki}  \& {Watanabe}}{{Shinnaka}
  et~al.}{2018}]{2018ApJ...864L..33S}
{Shinnaka} Y.,  {Kasuga} T.,  {Furusho} R.,  {Boice} D.~C.,  {Terai} T.,
  {Noda} H.,  {Namiki} N.,   {Watanabe} J.-i.,  2018, \mn@doi [\apjl]
  {10.3847/2041-8213/aadb3d}, \href
  {https://ui.adsabs.harvard.edu/abs/2018ApJ...864L..33S} {864, L33}

\bibitem[\protect\citeauthoryear{{Shkuratov} \& {Opanasenko}}{{Shkuratov} \&
  {Opanasenko}}{1992}]{1992Icar...99..468S}
{Shkuratov} I.~G.,  {Opanasenko} N.~V.,  1992, \mn@doi [\icarus]
  {10.1016/0019-1035(92)90161-Y}, \href
  {https://ui.adsabs.harvard.edu/abs/1992Icar...99..468S} {99, 468}

\bibitem[\protect\citeauthoryear{{Tabeshian}, {Wiegert}, {Ye}, {Hui}, {Gao}  \&
  {Tan}}{{Tabeshian} et~al.}{2019}]{2019AJ....158...30T}
{Tabeshian} M.,  {Wiegert} P.,  {Ye} Q.,  {Hui} M.-T.,  {Gao} X.,   {Tan} H.,
  2019, \mn@doi [\aj] {10.3847/1538-3881/ab245d}, \href
  {https://ui.adsabs.harvard.edu/abs/2019AJ....158...30T} {158, 30}

\bibitem[\protect\citeauthoryear{{Taylor} et~al.,}{{Taylor}
  et~al.}{2019}]{2019P&SS..167....1T}
{Taylor} P.~A.,  et~al., 2019, \mn@doi [\planss] {10.1016/j.pss.2019.01.009},
  \href {https://ui.adsabs.harvard.edu/abs/2019P&SS..167....1T} {167, 1}

\bibitem[\protect\citeauthoryear{{Tholen}}{{Tholen}}{1984}]{1984PhDT.........3T}
{Tholen} D.~J.,  1984, PhD thesis, University of Arizona, Tucson

\bibitem[\protect\citeauthoryear{{Tonry} et~al.,}{{Tonry}
  et~al.}{2012}]{2012ApJ...750...99T}
{Tonry} J.~L.,  et~al., 2012, \mn@doi [\apj] {10.1088/0004-637X/750/2/99},
  \href {https://ui.adsabs.harvard.edu/abs/2012ApJ...750...99T} {750, 99}

\bibitem[\protect\citeauthoryear{{Umov}}{{Umov}}{1905}]{1905Umov}
{Umov} N.~A.,  1905, Phis. Zeits., 6, 674

\bibitem[\protect\citeauthoryear{Virtanen et~al.,}{Virtanen
  et~al.}{2020}]{2020SciPy-NMeth}
Virtanen P.,  et~al., 2020, \mn@doi [Nature Methods]
  {10.1038/s41592-019-0686-2}, \href {https://rdcu.be/b08Wh} {17, 261}

\bibitem[\protect\citeauthoryear{{Watanabe}, {Takahashi}, {Sato}, {Watanabe},
  {Fukuhara}, {Hamamoto}  \& {Ozaki}}{{Watanabe}
  et~al.}{2012}]{2012SPIE.8446E..2OW}
{Watanabe} M.,  {Takahashi} Y.,  {Sato} M.,  {Watanabe} S.,  {Fukuhara} T.,
  {Hamamoto} K.,   {Ozaki} A.,  2012, in {McLean} I.~S.,  {Ramsay} S.~K.,
  {Takami} H.,  eds,  Society of Photo-Optical Instrumentation Engineers (SPIE)
  Conference Series Vol. 8446, Ground-based and Airborne Instrumentation for
  Astronomy IV. p. 84462O, \mn@doi{10.1117/12.925292}

\bibitem[\protect\citeauthoryear{{Whipple}}{{Whipple}}{1983}]{1983IAUC.3881....1W}
{Whipple} F.~L.,  1983, \iaucirc, \href
  {https://ui.adsabs.harvard.edu/abs/1983IAUC.3881....1W} {3881}

\bibitem[\protect\citeauthoryear{{Widorn}}{{Widorn}}{1967}]{1967AnWiD..27..109W}
{Widorn} T.,  1967, Annalen der Universitaets-Sternwarte Wien, Dritter Folge,
  \href {http://adsabs.harvard.edu/abs/1967AnWiD..27..109W} {27, 109}

\bibitem[\protect\citeauthoryear{{Worms}, {Renard}, {Hadamcik},
  {Levasseur-Regourd}  \& {Gayet}}{{Worms} et~al.}{1999}]{1999Icar..142..281W}
{Worms} J.-C.,  {Renard} J.-B.,  {Hadamcik} E.,  {Levasseur-Regourd} A.-C.,
  {Gayet} J.-F.,  1999, \mn@doi [\icarus] {10.1006/icar.1999.6188}, \href
  {https://ui.adsabs.harvard.edu/abs/1999Icar..142..281W} {142, 281}

\bibitem[\protect\citeauthoryear{{Yomogida} \& {Matsui}}{{Yomogida} \&
  {Matsui}}{1984}]{1984E&PSL..68...34Y}
{Yomogida} K.,  {Matsui} T.,  1984, \mn@doi [Earth and Planet. Sci. Lett.]
  {10.1016/0012-821X(84)90138-9}, \href
  {https://ui.adsabs.harvard.edu/abs/1984E&PSL..68...34Y} {68, 34}

\bibitem[\protect\citeauthoryear{{Zechmeister} \& {K{\"u}rster}}{{Zechmeister}
  \& {K{\"u}rster}}{2009}]{2009A&A...496..577Z}
{Zechmeister} M.,  {K{\"u}rster} M.,  2009, \mn@doi [\aap]
  {10.1051/0004-6361:200811296}, \href
  {https://ui.adsabs.harvard.edu/abs/2009A&A...496..577Z} {496, 577}

\bibitem[\protect\citeauthoryear{{Zellner}, {Leake}, {Lebertre}, {Duseaux}  \&
  {Dollfus}}{{Zellner} et~al.}{1977}]{1977LPSC....8.1091Z}
{Zellner} B.,  {Leake} M.,  {Lebertre} T.,  {Duseaux} M.,   {Dollfus} A.,
  1977, Lunar and Planet. Sci. Conf., \href
  {https://ui.adsabs.harvard.edu/abs/1977LPSC....8.1091Z} {1, 1091}

\bibitem[\protect\citeauthoryear{{Zimmerman}, {Farrell}, {Hartzell}, {Wang},
  {Horanyi}, {Hurley}  \& {Hibbitts}}{{Zimmerman}
  et~al.}{2016}]{2016JGRE..121.2150Z}
{Zimmerman} M.~I.,  {Farrell} W.~M.,  {Hartzell} C.~M.,  {Wang} X.,  {Horanyi}
  M.,  {Hurley} D.~M.,   {Hibbitts} K.,  2016, \mn@doi [\jgr]
  {10.1002/2016JE005049}, \href
  {https://ui.adsabs.harvard.edu/abs/2016JGRE..121.2150Z} {121, 2150}

\bibitem[\protect\citeauthoryear{{de Le{\'o}n}, {Pinilla-Alonso}, {Campins},
  {Licand ro}  \& {Marzo}}{{de Le{\'o}n} et~al.}{2012}]{2012Icar..218..196D}
{de Le{\'o}n} J.,  {Pinilla-Alonso} N.,  {Campins} H.,  {Licand ro} J.,
  {Marzo} G.~A.,  2012, \mn@doi [\icarus] {10.1016/j.icarus.2011.11.024}, \href
  {https://ui.adsabs.harvard.edu/abs/2012Icar..218..196D} {218, 196}

\bibitem[\protect\citeauthoryear{{van Dokkum}}{{van
  Dokkum}}{2001}]{2001PASP..113.1420V}
{van Dokkum} P.~G.,  2001, \mn@doi [\pasp] {10.1086/323894}, \href
  {https://ui.adsabs.harvard.edu/abs/2001PASP..113.1420V} {113, 1420}

\makeatother
\end{thebibliography}




\appendix

\section{NOT image reduction for subtracting stars}\label{app: NOT}

We usually discarded the MSI images when the asteroid was close (within 3$\times$FWHM) to the field stars. However, we could not discard the NOT images at large phase angles because of the lack of exposures. Especially in the NOT data on 2018 September 19, the asteroid was frequently contaminated by the field stars not only because these NOT data were taken without a field mask for polarimetry but also because the asteroid was close to the galactic plane. We also noticed that glares from very bright stars make aperture photometry difficult due to the severe sky gradients from the lights. To make the best use of the NOT data at $\alpha \geq  87.74 \degr$, we applied the following steps to data on 2018 September 12 and 19 to eliminate the influence of the field stars.

\begin{figure}
\includegraphics[width=\columnwidth]{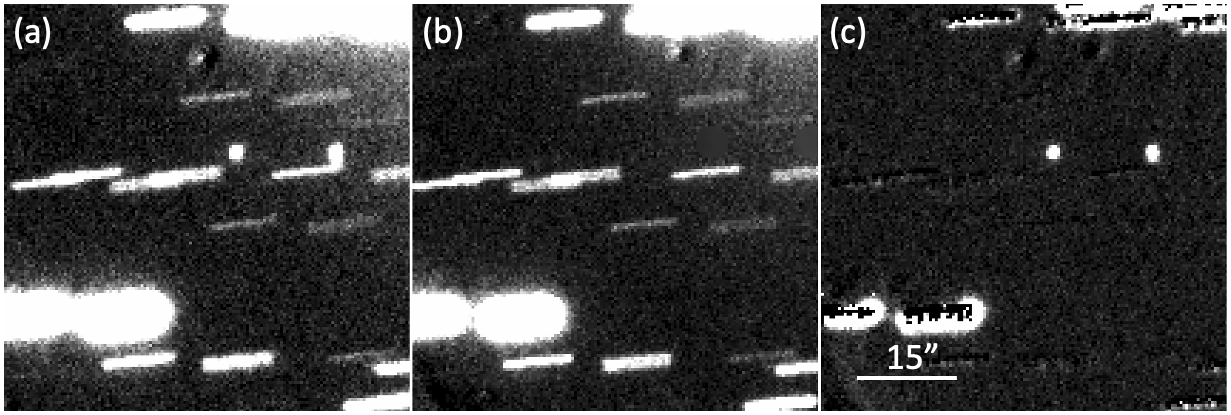}
\caption{Example images for the star subtraction. (a) an original image, (b) an image with field stars and without the asteroid, and (c) an image after field stars subtracted.}
\label{fig:starsubtraction}
\end{figure}

First, we identified the locations of field stars in both ordinary and extraordinary components using the Gaia star catalogue. The locations of stars brighter than 20.8 mags were specified on the CCD frame. Second, in each image taken in succession, we specified the locations of the asteroid in both ordinary and extraordinary components referring to an ephemeris and masked the pixel data within 3$\times$FWHM from the asteroid photocenter (Figure \ref{fig:starsubtraction} (b)). 
Because the asteroid moved to the field stars, we created images where field stars and sky background signals are recorded while the asteroid is not. Then, a set of two successive images (with the retarder angle of 0\degr and 45\degr or 22.5\degr and 67.5\degr) are co-added to match the star's positions, excluding the masked region for the asteroid. Finally, the original images (Figure \ref{fig:starsubtraction} (a)) were subtracted using the images without the asteroids to obtain the images (Figure \ref{fig:starsubtraction} (b)) where the contaminations of field stars are eliminated (Figure \ref{fig:starsubtraction} (c)).

\section{Polarimetric Phase Curve Fitting} \label{app: fitting}
This appendix introduces the outlines of the least square (LS) and Monte Carlo (MC) simulations to obtain the polarimetric parameters used in this work. Out of nine polarimetric measurements obtained in this work (Table \ref{table:polvalues}), one (2018 September 29) was not used due to its large uncertainty. All other data points are assumed to follow Gaussian (normal) distribution with mean $ P_r $ and standard deviation $ \sigma P_r $ (Table \ref{table:polvalues}). The non-Gaussianity of optical polarimetric measurements is not considered. We used freely available packages including PyMC3 (\citealt{2016ascl.soft10016S}; version 3.8) with scipy (\citealt{2020SciPy-NMeth}) on Python 3.8 environment. 

The polarimetric phase curve in Eq (\ref{eq:LM}) works as desired (zeros at $ \alpha = 0\degr$, $ \alpha_0 $, and $180\degr$) only if both $c_1$ and $c_2$ are positive, i.e., the ``bound'' case. If this condition is freed, it is called the ``unbound'' case in this work. In the MC simulations below, the default settings of PyMC3 is used with 20,000 samples per chain with four chains. The initial guess of the parameters were $ (h,\, \alpha_0,\, c_1,\, c_2) = (0.1 \,\mathrm{\% \,deg}^{-1}, 20 \mathrm{\degr},\, 0.1,\, 0.001) $. The identical MC simulations was done for the bound and unbound cases. 

\begin{figure}
\includegraphics[width=\columnwidth]{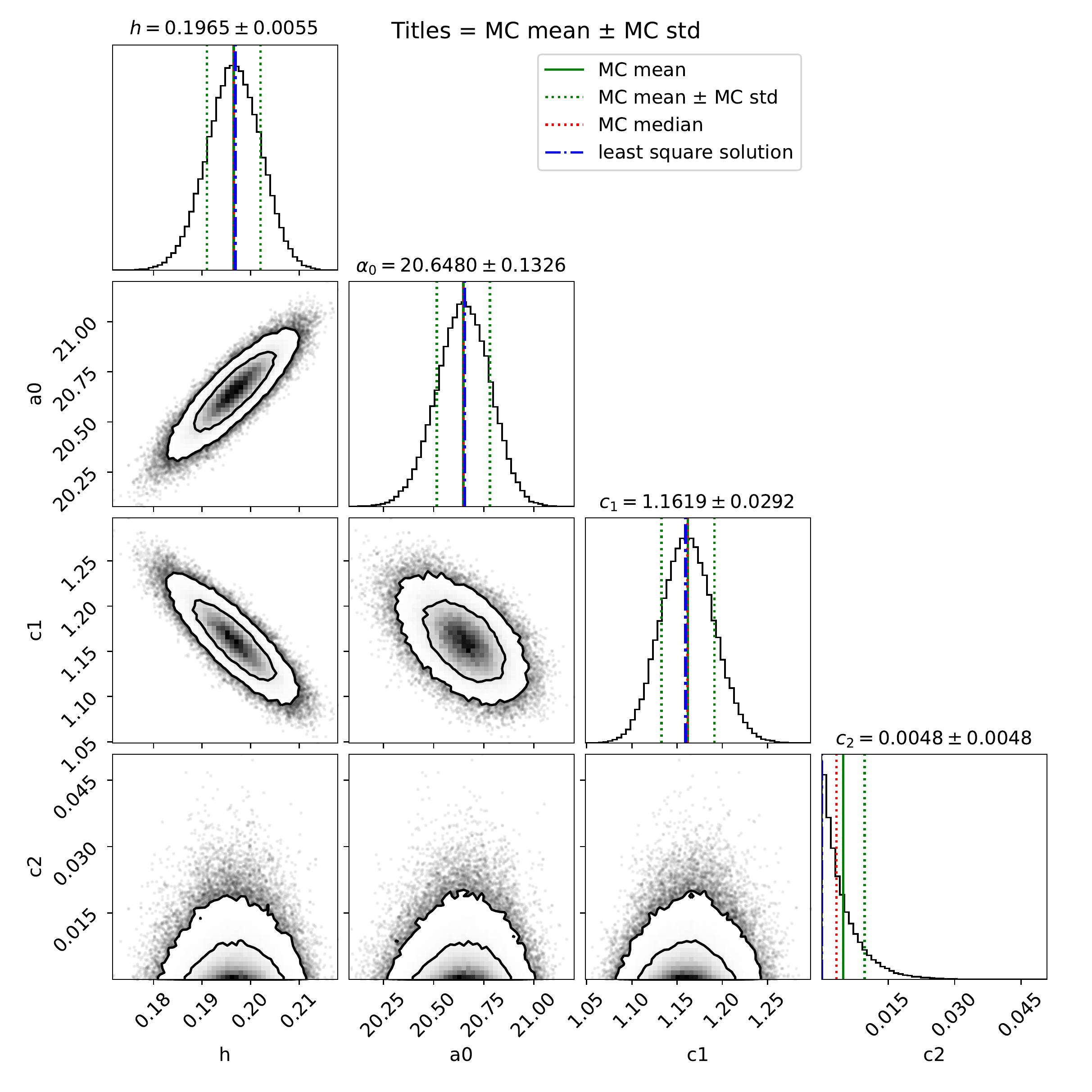}
\caption{The parameter estimation pair plots for the bound case (i.e., $c_1$ and $c_2$ are forced to be positive). The titles for each subplot gives the sample means and standard deviations from the Monte Carlo trace. In the posterior plots (diagonal panels), Monte Carlo (MC) mean and median are indicated as green solid and red dotted lines, respectively, although they are barely distinguishable. The mean $ \pm $ standard deviation is shown as green dotted lines. The blue dot-dashed lines show the least-square, i.e., the maximum likelihood estimation, which must be similar to the MC results.}
\label{fig:corner}
\end{figure}

For the bound case, uniform priors with the range $ h \in [0 \,\% \,\mathrm{deg}^{-1},\, 1 \,\% \,\mathrm{deg}^{-1}]$, $ \alpha_0 \in [10\degr,\, 35\degr]$, $c_1 \in [10^{-6},\, 3]$, and $c_2 \in [10^{-6},\, 3]$ are employed. The resulting parameter pair plots are shown in Figure \ref{fig:corner}. As visible, the posterior of $c_2$ is truncated at zero. For the slope $h$ and inversion angle $ \alpha_0 $, the MC means and medians match the LS estimations within an interval much less than the standard deviation. Other derived polarimetric parameters ($ P_\mathrm{min} $ and $ \alpha_\mathrm{min} $) are calculated for each MC sample by finding the minimum function value and its location. Then the sample mean and standard deviations of these were calculated, similar to all other parameters. 

After the MC samples are retrieved, the usual $ \chi^2 $ is calculated for each of those MC samples by
\begin{equation}
  \chi^2 = \sum_{i} \left ( \frac{P_\mathrm{r}^{\mathrm{(obs)}_i}(\alpha_i) - P_\mathrm{r}(\alpha_i; h,\, c_1,\, c_2,\, \alpha_0)}{\sigma P_\mathrm{r}^{\mathrm{(obs)}_i}} \right )^2 ~.
\end{equation}
The subscript $ i $ denotes each observation, $ P_\mathrm{r}^{\mathrm{(obs)}_i}(\alpha_i) $ and $ P_\mathrm{r}(\alpha_i) $ are the observed and model polarization degree, respectively, and $ \sigma P_\mathrm{r}^{\mathrm{(obs)}_i} $ is the Gaussian error-bar of the $ i $-th observation. If $ \chi^2_\mathrm{min} $ is the minimum $ \chi^2 $ among the MC samples, the code finds all other MC samples with $ \chi^2(h,\, c_1,\, c_2,\, \alpha_0) < \chi^2_\mathrm{min} + \Delta(\nu, \beta) $. Here,  $ \Delta$ is the inverse cumulative distribution function of the $ \chi^2 $ distribution, $ \beta $ is the significance level ($ \beta = 0.6827$ for 1-$ \sigma $), and $ \nu$ is the number of free parameters \citep[See, e.g., \S 15.6 of][]{press2007numerical}. Then each sample with this small $ \chi^2 $ value is the parameter set within the 1-sigma level confidence interval. The minimum and maximum of parameters $ (h,\, c_1,\, c_2,\, \alpha_0)$ are the 1-$ \sigma $ lower and upper bounds.

\begin{figure}
\includegraphics[width=\columnwidth]{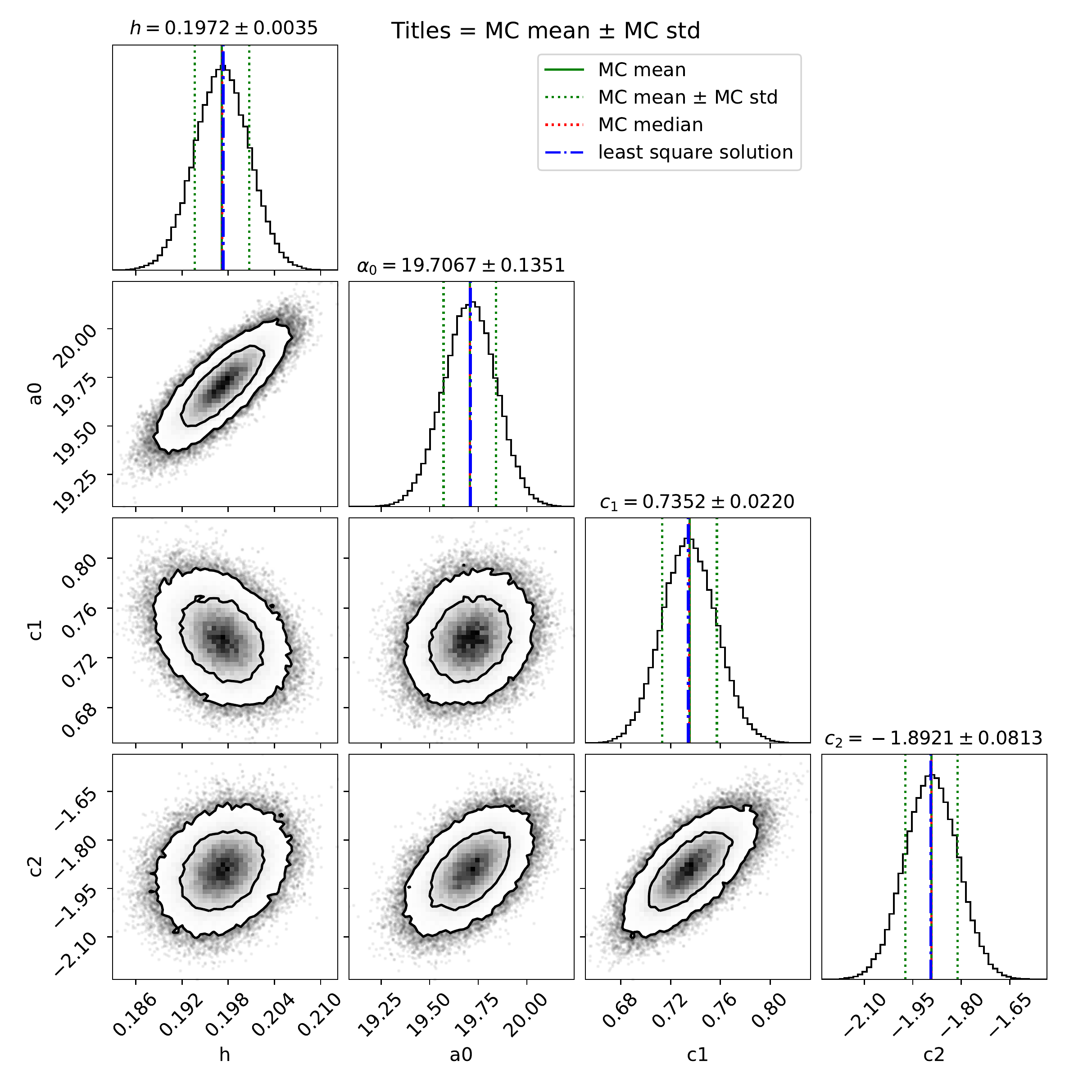}
\caption{Same as Figure \ref{fig:corner}, but for the unbound case (i.e., $c_1$ and $c_2$ are free to be negative).}
\label{fig:corner-unbound}
\end{figure}

For the unbound case, the prior is loosen: $c_1,\, c_2 \in [-1,\, 1]$. Similar pair plots are shown in Fig. \ref{fig:corner-unbound}. The negative values of $c_1$ and $c_2$ do not guarantee the $ P_\mathrm{r} = 0\,\%$ at $ \alpha = 0\degr$ and $180 \degr$, and the $ P_\mathrm{max} $ can even exceed $ 100\,\% $. This peculiar feature is visible in Fig. \ref{fig:bestfit}. 

It is clear from Fig. \ref{fig:bestfit} that the $ P_\mathrm{max} $ is significantly underestimated in the bound case. It is checked that increasing the uncertainties of the data points from \cite{2020PSJ.....1...15D} (mostly at small $ \alpha $) by a factor of 5 to 10 did not change the fitting results. This implies that the strong weighting to the small-uncertainty data points at small $ \alpha $ is not the main cause of the unsatisfactory fitting results near $ \alpha_\mathrm{max} $. 

Due to the random nature of MC simulation, MC mean and standard deviation values may change in every run but must reside within intervals much less than the nominal uncertainties. Other MC uncertainty measures, such as quantiles or the highest posterior density intervals, do not change our logic in this work (see the codes in DATA AVAILABILITY).

\bsp	
\label{lastpage}
\end{document}